\documentclass[usenatbib]{mn2e}
\usepackage{graphicx,times}
\usepackage{bm}
\usepackage{epsf}
\usepackage{amsmath}
\usepackage{amssymb}
\date{\today,~ $ $Revision: 0.9 $ $}
\def\la{\langle}
\def\ra{\rangle}
\def\n{\noindent}
\def\be{\begin{equation}}
\def\ee{\end{equation}}
\def\ben{\begin{eqnarray}}
\def\een{\end{eqnarray}}

\def\oh{\hat\Omega}

\def\bk{{\bf k}}

\def\inc{{\int_{r_i}^{r_q}}}

\def\bk{{\bf k}}

\def\bl{{\bf l}}
\def\bx{{\bf x}}
\def\2p{{(2\pi)^2}}

\def\bl{{\bf l}}
\def\be{\begin{equation}}
\def\ee{\end{equation}}
\def\beq{\begin{equation}}
\def\eeq{\end{equation}}
\def\ben{\begin{eqnarray}}
\def\een{\end{eqnarray}}

\def\oh{{\hat\Omega}}

\newcommand{\beqa}{\begin{eqnarray}}
\newcommand{\eeqa}{\end{eqnarray}}

\def\ikap0{{\cal J}_{\theta_0}(r)}

\def\av{\langle \myF_{\alpha}^2\rangle}

\def\one1{\langle \myF_{(i)}\myF_{(j)} \rangle}
\def\one{{[\bar \xi^{(ij)}]}}

\def\myF{{\delta\cal F}}
\def\myJ{{\cal J}}
\begin{document}
\onecolumn
\title[The Statistics of Cosmological Lyman-$\alpha$ Absorption]
{The Statistics of Cosmological Lyman-$\alpha$ Absorption}
\author[Munshi et al.]
{Dipak Munshi$^{1}$, Peter Coles$^{1}$, Matteo Viel$^{2,3}$\\ 
$^{1}$School of Physics and Astronomy, Cardiff University, Queen's
Buildings, 5 The Parade, Cardiff, CF24 3AA, UK, \\
$^{2}$ INAF-Observatorio Astronomico di Trieste, Via G.B. Tiepolo 11, I-34131, Trieste, Italy,\\
$^{3}$  INFN sez. Trieste, via Valerio 2, 34127, Trieste, Italy.}
\maketitle
\begin{abstract}
We study the effect of the non-Gaussianity induced by gravitational evolution upon the
statistical properties of absorption in quasar (QSO) spectra.  
Using the generic hierarchical ansatz and the lognormal approximation 
we derive the analytical expressions for the one-point
PDF as well as for the joint two-point probability distribution (2PDF)
of transmitted fluxes in two neighbouring QSOs. These flux PDFs are
constructed in 3D as well as in projection (i.e. in 2D).  The PDFs are
constructed by relating the lower-order moments, i.e. cumulants and
cumulant correlators, of the fluxes to the 3D neutral
hydrogen distribution which is, in turn, expressed as a function of the
underlying dark matter distribution. The lower-order moments are next
modelled using a generating function formalism in the context of a
{\em minimal tree-model} for the higher-order correlation hierarchy.
These different approximations give nearly identical results for the
range of redshifts probed, and we also find a very good agreement
between our predictions and outputs of hydrodynamical simulations. The
formalism developed here for the joint statistics of flux-decrements
concerning two lines of sight can be extended to multiple
lines of sight, which could be particularly important for the 3D
reconstruction of the cosmic web from QSO spectra (e.g. in the BOSS
survey). These statistics probe the underlying projected neutral
hydrogen field and are thus linked to ``hot-spots'' of absorption.  The
results for the PDF and the bias presented here use the same
functional forms of scaling functions that have previously been employed
for the modelling of other cosmological observation such as the
Sunyaev-Zel'dovich effect.
\end{abstract}
\section{Introduction}
Ongoing Cosmic Microwave Background (CMB) experiments such as
Planck\footnote{http://www.rssd.esa.int/Planck}, the Atcama-Cosmology
Telescope\footnote{http://www.physics.princeton.edu/act/} (ACT) and
the South Pole Telescope\footnote{http://pole.uchicago.edu/} (SPT)
will pinpoint the cosmological parameters that describe the background
geometry and dynamics of the Universe in an unprecedented
detail. Along with very precise constraints on the structure of the
Universe on the largest scales, smaller-scales observables will be
crucial in order to further constrain the cosmological concordance
model or find possible deviations from it.  In particular, galaxy
clustering (BOSS, 6dF, etc.) surveys and future weak lensing and
clustering observations (e.g. EUCLID) could probe smaller scales and
new redshif regimes.  Spectroscopic surveys such as
BOSS\footnote{http://cosmology.lbl.gov/BOSS/} (and BIG BOSS) will also
trace the large scale distribution of the baryonic matter in the
Universe through the study of the flux distribution of the
Lyman-${\alpha}$ absorption systems in a very large number of quasar
(QSO) spectra.

The Lyman-$\alpha$ ``forest'', the many absorption features in QSO
spectra, produced by intervening neutral hydrogen in the intergalactic
medium (IGM) along the line-of-sight, is well known to be an important
cosmological probe (for a recent review see Meiksin 2009). In the
standard cosmological paradigm, the IGM consists of mildly non-linear
gas, making up the cosmic web, that traces the dark matter and is
photo-heated by a Ultra Violet (UV)-background.  The Lyman-$\alpha$
forest is thus the main probe of the IGM and it has been shown to
arise naturally in hierarchical structure formation scenarios.
Astrophysical effects produced by feedback from galaxies and/or AGNs
do not seem to strongly affect the vast majority of the baryons in the
cosmic web \citep{Mc05,T02}, thereby this can be used as a dark matter
tracer.  The relation between the Lyman-${\alpha}$ forest flux and the
underlying matter field is a nonlinear one and it is generally
expected that statistics of Lyman-$\alpha$ are biased relative to the
underlying dark matter distribution. The Lyman-${\alpha}$ forest has
been studied using variety of analytical techniques such as the
Zel'dovich approximation, which is valid in the quasilinear regime and
often used in modelling of the nonlinear gravitational clustering
\citep{DS97, HGZ97,McGill90,MM02}.  In addition, the lognormal
approximation \citep{CJ91} is frequently used to model the statistics
of Lyman-$\alpha$ forest \citep{Bi93,GH96,BD97,RPT01,Viel02}.  Models
based on the hierarchical or scaling {\em ansatz}
\citep{BS89,BeS92,BS99} for higher-order correlation functions have
also been investigated in order to model the statistics of
Lyman-$\alpha$ forest \citep{VSS99}.
  
In addition to analytical modelling, hydrodynamical simulations have
also played a very important role in this field
(e.g. \cite{Cen94,GH98,Cr98,Cr99,MW01}) and support the simple
analytical picture. Thus, analytical schemes, including the ones that
we develop here, can be calibrated using numerical simulations and
this in turn allows to explore a large parameter space efficiently.
However, numerical simulations are required to resolve the Jeans scale
of the photo-ionized warm IGM, and this requirement typically means
small box sizes that sample larger scales modes rather poorly.  Indeed
several semi-numerical prescriptions, which are not entirely based on
hydrodynamical simulations, have also been developed to model
Lyman-$\alpha$ flux and recover the correlation function from observed
data sets \citep{Sls11}.
\begin{figure}
\begin{center}
{\epsfxsize=10 cm \epsfysize=5.1 cm {\epsfbox[29 436 581 714]{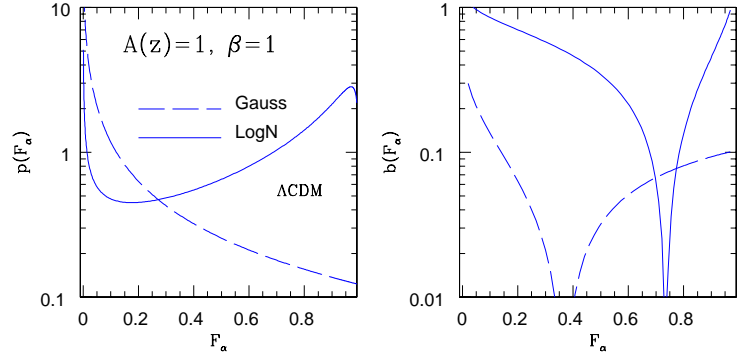 }}}
\end{center}
\caption{The PDF $p(F_{\alpha})$ and bias $b(F_{\alpha})$ of the flux $F_{\alpha}$ is plotted as a function of the flux $F_{\alpha}$. 
The PDF (left-panel) and the bias (right-panel) both are constructed 
using a lognormal model for the underlying mass distribution. The resulting PDFs for $\tilde\delta_{}$ are
next transformed into the flux PDFs using the fluctuating Gunn-Peterson approximation Eq.(\ref{eq:gp}).
We compare the results from the lognormal approximation (solid lines) against the one based on the Gaussian approximation (dashed lines).
The values of $A(z)$ and $\beta$ are constructed using the functional fit in Eq.(\ref{eq:fit}) given by \citep{Kim07} see text for more details.
The bias changes signs at an intermediate flux value which depends on cosmology and hydrodynamical parameters that define the
equation of state of the photo-ionized medium $A(z)$ and $\beta$. A fiducial value for the variance $\sigma=3$ was assumed for this plot. }
\label{fig:demo}
\end{figure}

The two most commonly used approaches in Lyman-${\alpha}$ studies are based either on
decomposing the information encoded in the transmitted flux via Voigt profile fitting
or treating the flux as a continuous field. 

In the first approach, the shapes and clustering properties of
absorption lines fitted by Voigt profiles have been investigated in
variety of studies involving the temperature of the IGM
\citep{STLE99,MMR01}, in order to constrain the reionization history
\citep{TSS02,HH03}, to measure the matter power spectrum and
cosmological parameters
\citep{Croft99,VHS04,MSB06,MM99,Rol03,Copp06,Gui07,VH06}.

By using the second set of methods, statistical properties of the flux
such as the mean flux level, flux PDF, flux power spectrum
\citep{VHS04,Viel08,SSM06} and flux bi-spectrum are typically employed
to explore flux statistics. For example, it has been shown that the
mean-flux level can be used to constrain the amplitude of
intergalactic UV background \citep{T04,Bolt05} while the flux PDF
\citep{Mc09, Bolt09} is sensitive to the thermal evolution of the IGM
(see also \cite{L06}). The flux power spectrum on the other hand
can be used to constrain the cosmological parameters and nature of
dark matter \citep{Croft02}.  The flux-bispectrum
\citep{Viel09} contains useful information about the primordial, as
well as gravity-induced (i.e. secondary) non-Gaussianity. The data
typically used in these investigations consists mainly of two
different sets of QSO spectra: the SDSS low resolution low
signal-to-noise spectra and UVES/VLT or HIRES/KECK high resolution
spectra. The number of SDSS spectra is about a factor $\sim 200$
larger than that of high resolution samples though the later probes
the smaller scales with greater accuracy.

More recently, it has been argued that BOSS-like QSO spectroscopic
surveys could detect Baryon Acoustic Oscillations (BAO) signatures at
high redshift \citep{ME07} and a sample of QSO pairs
can constrain the geometry of the high redshift universe \citep{Mc03}. Furthermore, analysis of coincident absorption lines in QSO
pairs can also allow departures from the Hubble flow and
non-gravitational effects to be measured \citep{Rauch05}.

The SDSS-III/BOSS survey aims at identifying and observing more than
160,000 QSO over $\sim 10,000$ square degrees within a redshift range
$z=(2.15-3.5)$. This survey is primarily design to studying baryonic
acoustic oscillations by performing a full 3D sampling of the matter
density.  Such studies will also provide an unprecedented opportunity
to study the clustering statistics using projected Lyman-$\alpha$ flux
decrements of QSOs.

In this paper, we will consider the Lyman-$\alpha$ flux decrement in
two dimensions (2D) which is related the projected density of neutral hydrogen. The
statistical study of projected Lyman-$\alpha$ flux decrement can be
performed using one- or two-point PDF or their lower order moments.
The PDFs contain information of cumulants or their correlators to an
arbitrary order and can be constructed using well-established
machinery of hierarchical ansatz \citep{MCM1,MCM2,MCM3}. Several
authors have recently studied the lower-order cumulants of
Lyman-$\alpha$ forests and cross-correlated them against weak lensing
convergence as well as to the CMB sky \citep{VDSS09,VVDS11}.  The
results presented here are complementary to such studies as we take
into account the lower order moments to an arbitrary order not just
for the one-point cumulants but also for their two-point counterparts
or {\em cumulant correlators}. The cumulant correlators are the
two-point analogues of one-point cumulants and are already in use in
different areas of cosmology, e.g.  in analysis of galaxy surveys
\citep{SS97}. The lowest in the two-point hierarchy is the two-point
correlation function. In the context of Lyman-$\alpha$ studies, the
two-point correlation function has already been introduced in studies
involving two neighbouring line of sights \citep{Viel02,Dod02}.  Our study
generalizes these results to probe non-Gaussian correlation functions
involving multiple line of sight. We model the statistics of
underlying neutral hydrogen distribution using lognormal distribution
as well as an extension of perturbation theory approach \citep{VaMu04}
in 3D.  Predictions from these models are then tested using
hydrodynamical simulations at three different redshifts
($z=2,3,4$). Next we use these results to build and test statistical
description of the projected flux distribution. The results presented
here can be generalized to cross-correlation studies involving
external data sets and Lyman-$\alpha$ flux distribution. The correlation functions (CCs) are
equivalent to their Fourier (harmonic) space counterpart, the
multispectra, recently introduced by \citep{MuHe09}.
 
The plan of this paper is as follows. In \textsection\ref{sec:form} we
introduce the notations and define the relevant quantities such as the
transmitted flux and its relation to the underlying density
contrast. In \textsection\ref{sec:sims} we give details of the
simulations that were used in our study.  In \textsection\ref{sec:3D}
details of modelling of the 3D flux are presented.  In
\textsection\ref{sec:lower} we derive the lower order statistics for
the flux in terms of that of the underlying density contrast. In
\textsection\ref{sec:hier} we provide a very brief introduction to the
hierarchical ansatz which we use to model the statistics of underlying
density contrast.  In \textsection\ref{sec:pdf} we provide derivation
of the PDF and the bias associated with the flux distribution and
finally \textsection\ref{sec:conclu} is left for discussion of
results.  We also provide a brief appendix introducing the lognormal
approximation as well as the hierarchical ansatz.
\section{Notation}
\label{sec:form}
 We will be using the following form of the Robertson-Walker line element for the
background geometry of the universe:
\begin{equation}
ds^2 = -c^2 dt^2 + a^2(t)( dr^2 + d_A^2(r)(d\theta^2 + \sin^2\theta d\phi^2))
\end{equation}
\begin{figure}
\begin{center}
{\epsfxsize=12  cm \epsfysize=5.1 cm {\epsfbox[23 476 587 714]{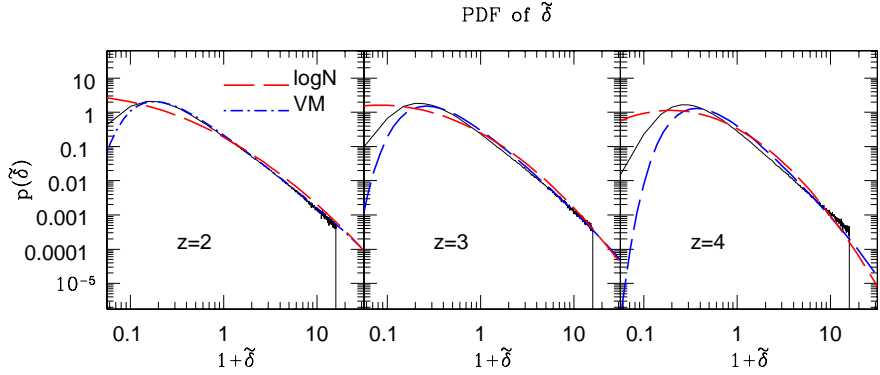}}}
\end{center}
\caption{The PDF of the density contrast $p(\tilde\delta_{})$ as a function of the density contrast $\tilde\delta_{}$.
The left ($z=2$), middle ($z=3$) and right ($z=4$) panels correspond the to different redshifts $z$ studied. Two different 
analytical models are displayed the lognormal approximation (long-dashed) and the model proposed by 
\citep{VaMu04} (short- and long-dashed) denoted as VM.The solid lines correspond to the results from simulations. Notice that
the analytical results and simulations for positive (over-dense) $\delta_{}$ values of the PDF agree extremely well for
all three redshifts investigated in this work. The PDF was estimated using $128^3$ grid and can resolve 
PDF as low as $O(10^{-6})$. However the PDF becomes increasingly dominated by the presence or absence of rare large overdensities.}
\label{fig:delta_pdf}
\end{figure} 
\begin{figure}
\begin{center}
{\epsfxsize=12  cm \epsfysize=5.1 cm {\epsfbox[23 476 587 714]{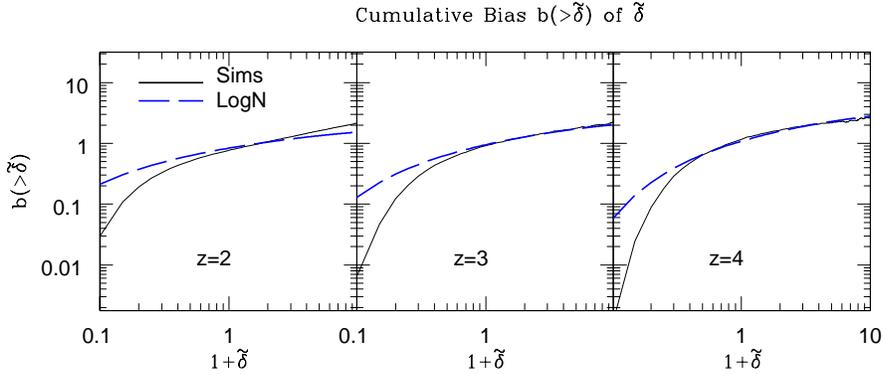}}}
\end{center}
\caption{The PDF of density contrast $p(\tilde\delta)$ as a function of the density contrast $\tilde\delta_{}$.
The left ($z=2$), middle ($z=3$) and right ($z=4$) panels correspond to different redshift $z$. Two different 
analytical models are displayed the lognormal approximation (long-dashed) and the model proposed by 
\citep{VaMu04} (short- and long-dashed) denoted as VM. The PDF is constructed on a grid of $128^3$. The lowest 
probability that can be estimated on this is grid is $O(10^{-6})$. However higher $\tilde\delta_{}$ tails of PDFs are increasingly
dominated by the presence (or absence) of rare over (under)dense objects.}
\label{fig:delta_bias}
\end{figure} 
The angular position on the surface of the sky is specified by the unit vector $\hat\Omega=(\theta,\phi)$.  The
scale factor of the Universe is given by $a(t)$.
We have denoted the comoving angular diameter distance by $d_A(r)$ where $r$ denotes the
comoving radial distance to redshift $z$; $d_A(r)= {\rm K}^{-1/2}\sin
({\rm K}^{1/2} r)$ for positive curvature, $d_A(r) = (-{\rm
K})^{-1/2}\sinh ((-{\rm K})^{1/2}r)$ for negative curvature and $r$
for the flat universe. For a present value of  of ${\rm H}_0$ and
$\Omega_{\rm M}$ we have ${\rm K}= (\Omega_{\rm M} -1){\rm H}_0^2$. The Hubble constant is denoted by ${\rm H}_0$. 
The comoving radial
distance $r(z)$ at a redshift $z$ can be expressed through the following LOS integration, with $z'$ playing
the role of intermediate redshift along the line of sight:
\be
r(z)= \int_0^z {dz' \over {{\rm H}_0 \sqrt{\Omega_{\rm M}(1+z')^3 + \Omega_{\rm K}(1+z')^2 + \Omega_{\Lambda}} }}.
\ee
Throughout, in our calculation, we will adopt 
${\rm H}_0=100h\;{\rm kms}^{-1}{\rm Mpc}^{-1}$ with $h=0.7$ and $\sigma_8=0.87$. Here $\Omega_{\rm M}$, $\Omega_{\Lambda}$ and $\Omega_{\rm K}$ are dark-matter,
vacuume-energy and curvature contribution to cosmic density $\Omega_{\rm M}+\Omega_{\Lambda}+\Omega_{\rm K}=1$. For a flat Universe
$\Omega_{\rm K}=0$. 
\section{Simulations}
\label{sec:sims}
\begin{figure}
\begin{center}
{\epsfxsize=13 cm \epsfysize=8.15 cm {\epsfbox[23 23 656 365]{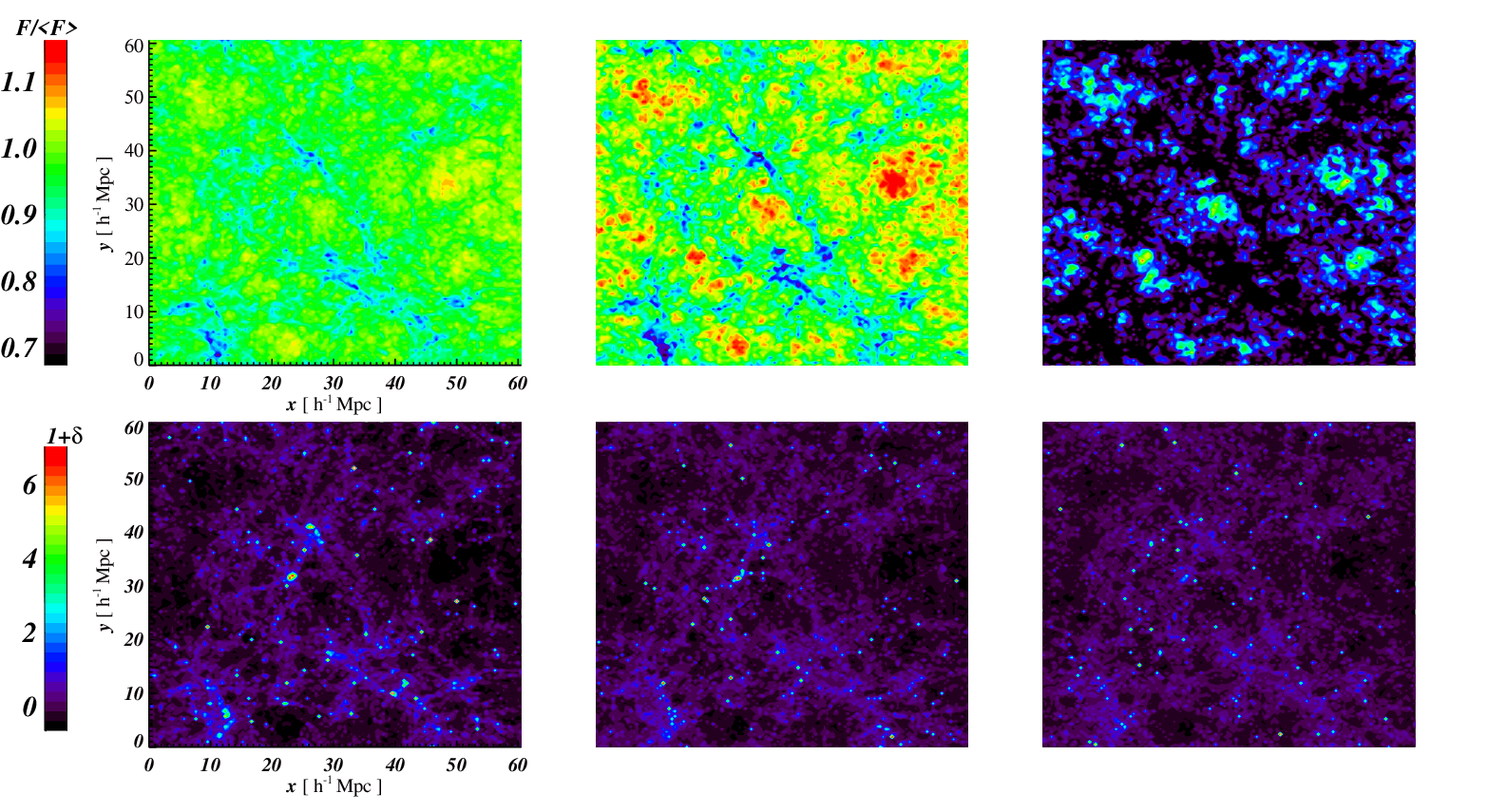}}}
\end{center}
\caption{We show the projected transmitted Lyman$-\alpha$ flux (top
  panels) and projected gas density (bottom panels) in the simulated
  volume at z=2,3,4 (left, middle and right columns, respectively).
  The gas density is extracted at $128^3$ grid points using a CIC or
  cloud-in-cell interpolation scheme: the values shown in the bottom
  panels are the projected $\tilde\delta$ along $128^2$ lines of sight
  (along the z-axis). The growth of the cosmic structures can be
  appreciated from high to low redshift.  In the upper panels we show
  the mean value $\la F_{\alpha}\ra$ along the same $128^2$ lines of sight, the
  values at the three redshift are all normalized to the $\la
  F_{\alpha}(z=2)\ra$ value and the spectra are simulated by using the
  exact definition of the transmitted flux and not
  approximations. Since the mean flux is a strongly evolving function
  of redshift, the growth of cosmic structure in Lyman-$\alpha$ flux
  is more difficult to interpret than in the corresponding density
  slices and simulations and/or models are needed. In the upper panels,
  voids correspond to regions of high-transmissivity, while dense
  regions produce absorptions. At $z=2$ the $\la F_{\alpha}\ra $
  values fluctuate by about 10\% around the mean, while at $z=4$ this
  value becomes 20\% (since the mean transmitted flux is about two
  times smaller at this redshift).  The simulation refers to a
  hydrodynamical run with $2\times 512^3$ gas and DM particles in a
  $60^3$ (Mpc ${\rm h}^{-1})^3$ periodic comoving volume.}
\label{fig:flux_pdf}
\end{figure} 
We use simulations run with the parallel hydrodynamical (TreeSPH) code
{\small{GADGET-2}} based on the conservative `entropy-formulation' of
SPH \citep{Sp05}.  These consist of a cosmological volume with
periodic boundary conditions filled with an equal number of dark
matter and gas particles.  Radiative cooling and heating processes
were followed for a primordial mix of hydrogen and helium.  We assumed
a mean Ultraviolet Background similar to that proposed by
\cite{HM96} produced by quasars and galaxies as given by with
helium heating rates multiplied by a factor 3.3 in order to better fit
observational constraints on the temperature evolution of the IGM.
This background gives a hydrogen ionisation rate $\Gamma_{12}\sim 1$
at the redshifts $z=2-4$ of interest here \citep{Bolt05}.  The star
formation criterion is a very simple, one that converts 
all the gas particles into stars whose temperature falls
below $10^5$ K and whose density contrast is larger than 1000 (it has
been shown that the star formation criterion has a negligible impact
on flux statistics).  More details can be found in
\citep{VHS04}.

The cosmological reference model corresponds to a `fiducial'
$\Lambda$CDM Universe with parameters, at $z=0$, $\Omega_{\rm m
}=0.3,\ \Omega_{\rm \Lambda}=0.7,\ \Omega_{\rm b }=0.05$, $n_{\rm
  s}=1$, and ${\rm H}_0 = 70$ km s$^{-1}$ Mpc$^{-1}$ and $\sigma_8=0.85$.
The initial conditions are generated at $z=49$ using as linear matter
power the one extracted using the {\small CAMB} software.  We have
used $2\times 512^3$ dark matter and gas particles in a volume of
linear size $60\ {\rm h}^{-1}$ comoving Mpc${\rm h}^-1$.  The gravitational
softening was set to 4 ${\rm h}^{-1}$ kpc in comoving units for all
particles. The mass per particle is $1.1\times10^8 $M$_{\odot}\ {\rm h}^{-1}$ and
$1.8\times10^7 $M$_{\odot}\ {\rm h}^{-1}$ for DM and gas particles, respectively.
This simulation should have sufficient resolution to properly
reproduce most of flux statistics at least at $z\le 3$ and marginally
at $z=4$: the flux power should be in fact converged with this setup
and also (marginally) the flux PDF. However, since we are dealing with
projected quantities, we believe that numerical convergence issues are
less severe than for other small scale flux observables.

We select $128^2$ grid points in the $x-y$ plane and we interpolate
along lines of sights parallel to the $z-$axis using a Cloud-In-Cell (CIC)
algorithm. The Lyman-$\alpha$ flux is also computed in redshift space
along the same lines of sight by using the exact definition of transmitted flux
and not the Fluctuating Gunn-Peterson Approximation. Peculiar
velocities, neutral hydrogen fraction and gas temperature are
calculated along the lines of sight and a Voigt profile is used to obtain the
mock quasar spectra.  The density and transmitted flux are then
projected along the simulated lines of sight in the $z-$direction.
\section{The Lyman-$\alpha$ flux and the IGM in 3D}
\label{sec:3D}
The fluctuating Gunn-Peterson approximation \citep{GP65} allows us to relate the transmitted flux  
$F_{\alpha}(z,\oh)$ along a line of sight at a direction $\oh\equiv (\theta,\phi)$ at a redshift $z$ with the 
fluctuation in neutral hydrogen density contrast $\tilde \delta_{}$:
\ben
&& {F}_{\alpha}(z,\oh) = \exp\left [-A(z)(1+\tilde \delta_{}(z,\oh))^{\beta}\right]; \\
\label{eq:flux}
&& A(z)\equiv 0.0023 (1+z)^{3.65}; \quad\quad \beta \equiv 2 - 0.7(\gamma-1);\quad \gamma=1.3.
\label{eq:gp}
\een The parameters $A(z)$ and $\beta$ are two redshift-dependent
functions relating the flux fluctuations to the dark matter
overdensities.  The parameter $A$ is related to the level of mean
transmitted flux which is accurately measured (e.g.
\cite{Kim07}). It also depends on inputs from baryonic physics such as
the baryonic fraction, IGM temperature, photo-ionization rate as well
as background cosmological parameters.  The power-law index
$\gamma$ of the IGM temperature-density relation determimes the value of
the parameter $\beta$ and is relatively independent of redshift $z$
\citep{HG97,Kim07} (this is of course assuming fluctuations in
temperature due to e.g. reionization play a sub-dominant role).  The
allowed range of values for $\gamma$ is $\gamma = (0.7-1.5)$ we will
use $\gamma=1.3$ for our calculation, which is a good approximation
to the actual value of the hydrodynamical simulation.
\subsection{PDF and Bias in 3D}
We use the lognormal distribution to model the distribution of $\tilde\delta$;
the resulting flux PDF is obtained via a change of variable:
\be p(F_{\alpha}) = p(\tilde \delta_{})\left
|d\tilde \delta_{} \over dF_{\alpha} \right |; \quad\quad
p(F_{\alpha}^{(1)}, F_{\alpha}^{(2)}) = p(\tilde \delta_{1},
\tilde\delta^{}_{2})\left |d\tilde \delta^{}_{1} \over
dF^{(1)}_{\alpha} \right |\left |d\tilde \delta^{}_{2} \over
dF^{(2)}_{\alpha} \right |.
\label{eq:transform}
\ee
In our notation $\tilde\delta_{(i)}=\tilde\delta_{}(\oh_i,z_i)$ and a similar notation is adopted 
for the fluxes along different lines of sight $F^{(i)}_{\alpha}=F_{\alpha}(\oh_i,z_i)$. For 3D studies we have compared flux 
from neighbouring lines of sight at the same redshift. We have compared the PDFs using this approach
for three different redshifts $z=2,3,4$.  The PDFs for $p(\delta_{})$ are constructed using lognormal prescriptions (Eq.(\ref{eq:logn1}) 
and Eq.(\ref{eq:logn2a})).

The lognormal distribution is also used in the analysis based on the
fitted column density distribution of absorption lines.  In such
studies the lognormal distribution is introduced to relate
$\tilde\delta_{}$ and number density of neutral hydrogen $n_{\rm HI}$.
In relating the flux distribution and $\tilde\delta_{}$ on the other
hand we assume that $\tilde \delta_{}$ traces the underlying dark
matter distribution which can be described well by lognormal
distribution. The lognormal distribution here is being used to model
the effect of gravity induced non-linearity on PDFs of matter density
contrast $\delta$. A similar but somewhat different approach was used
by \cite{Viel02} who used lognormal approximation to map the density
field to flux. In our case we directly map the PDF of the $\tilde
\delta_{}$ to that of the flux using simply a change of variable
Eq.(\ref{eq:gp}). The construction of the PDF of $\tilde \delta_{}$ is
done using lognormal approximation as well as using extensions of
perturbative methods \citep{VaMu04}. These methods give near identical
results.  The hierarchical ansatz (to be introduced later; see
Appendix-B) will be used too for the construction of one-point and
two-point PDFs.  It is interesting to point out that it has also been used
by \citep{VSS99} to predict the clustering Lyman-$\alpha$ absorbers.
It was used to construct an unifying model from the clustering of low
column-density clouds to the collapsed dense damped systems.  The
methods that we present here is complementary to these studies as we
directly probe the statistics of the transmitted flux.
\subsection{Numerical Results in 3D}
In Figure (\ref{fig:delta_pdf}) we have plotted the PDF of $p(\tilde \delta_{})$ as a function of $\tilde \delta_{}$. The PDF was computed by binning the data
on a $128^3$ grid. This allows us to resolve the PDF down to $O(10^{-6})$. Different panels show the PDF for 
three different redshifts. We also show the results from different analytical calculations. The two-point PDF or 2PDF denotes the
joint PDF of $\tilde\delta_{}$ at two different points. Typically the 2PDF can be written as a product of two 1PDFs at these two-point
and additional correction terms. The corrective terms represents the contribution due to correlation. In general in the 
large separation limit when the two-point correlation is weak compared to the variance (i.e. $\xi_{12}^{\delta} <\bar\xi_2^{\delta} $; $\xi_{12}^{\delta} = \la \delta(\bx_1)\delta(\bx_2) \ra$ is the two-point correlation function and $\bar\xi_2$ is its volume average), the 2PDF takes the following form in 3D:
\be
p(\tilde\delta_1,\tilde\delta_2)d\tilde\delta_1d\tilde\delta_2= 
p(\tilde\delta_1)p(\tilde\delta_2)[1+ b(\tilde\delta_1)\xi^{\delta}_{12}b(\tilde\delta_2)]d\tilde\delta_1d\tilde\delta_2.
\ee
This equation describes the two-point PDF (or 2PDF) $p(\delta_1,\delta_2)$ of the underlying density contrast $\delta$ in terms of PDFs $p(\delta)$ and bias $b(\delta)$. 
An identical relation also holds for $\tilde\delta_{}$
above the Jeans scale. In general the (differential) bias $b(\eta)$ is difficult to estimate. To increase signal-to-noise
it is useful to compute the integrated bias beyond a threshold \citep{B94}:
\be
b(>\delta_0) \equiv \left [ \int_{\delta_0}^{\infty} p(\tilde\delta)b(\tilde\delta)d\tilde\delta \over 
\int_{\tilde\delta_0}^{\infty} p(\tilde\delta)d\tilde\delta \right ] =
{1 \over \sqrt {\xi^{\delta}_{12}}}\left [ {\int_{\delta_0}^{\infty} \int_{\delta_0}^{\infty} p(\tilde\delta_1,\tilde\delta_2) 
d\tilde\delta_1 d\tilde\delta_2 \over [\int_{\delta_0}^{\infty} p(\tilde\delta) d\tilde\delta ]^2} -1 \right ]^{1/2}.
\label{eq:int_bias}
\ee
For our numerical estimation of bias associated with the $\tilde\delta_{}$ we have used Eq.(\ref{eq:int_bias}) to 
compute the cumulative bias associated with $\tilde\delta_{}$. In Figure (\ref{fig:delta_bias}) we have presented the 
numerical results for $b(>\delta_{0})$ against the theoretical expectations from lognoraml distribution.
As expected the lognormal distribution is quite accurate in prediction of the numerical results for over dense regions.
We show results for three different redshift relevant to high redshift Lyman-$\alpha$ studies. Notice that
the high $\tilde\delta_{}$ tail of the PDF is reproduced well by the lognormal approximation Eq.(\ref{eq:log_bias}). The departure 
at lower $\tilde\delta_{}$ is partly related to the smoothing introduced in hydrodynamical simulation during interpolation.
Indeed we only use the leading order term $\xi^{\delta}_{12}/\bar\xi^{\delta}$ in our calculation of bias from
the lognormal distribution. These results are valid in the large separation limit. Additional corrective terms 
can be included if necessary.

\begin{figure}
\begin{center}
{\epsfxsize=13 cm \epsfysize=5.15 cm {\epsfbox[23 476 587 714]{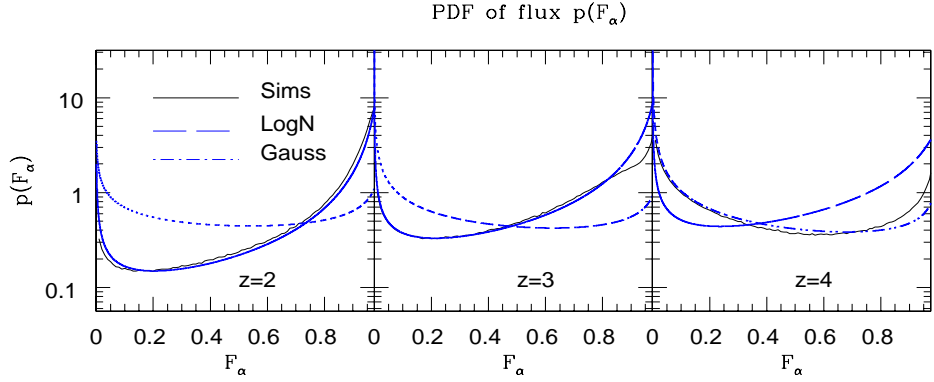}}}
\end{center}
\caption{The PDF of the flux $p(F_{\alpha})$ as a function of the flux $F_{\alpha}$.
The left ($z=2$), middle ($z=3$) and right ($z=4$) panels correspond to different redshift $z$. Two different 
analytical models are displayed the lognormal approximation (long-dashed) and the Gaussian approximation 
(short and long-dashed). The flux PDF is constructed from the PDF of $\tilde\delta_{}$ using Eq.(\ref{eq:gp}).
The results from the simulation are also depicted (solid lines).}
\label{fig:3Dflux_pdf}
\end{figure}

\begin{figure}
\begin{center}
{\epsfxsize=13 cm \epsfysize=5.15 cm {\epsfbox[23 476 587 714]{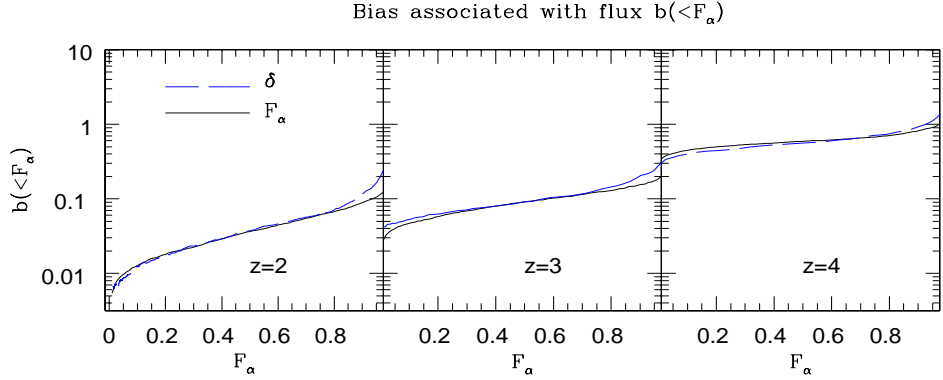}}}
\end{center}
\caption{The bias of the flux $b(<F_{\alpha})$ as a function of the flux $F_{\alpha}$.
The left ($z=2$), middle ($z=3$) and right ($z=4$) panels correspond to different redshift $z$.
The solid lines correspond to direct estimates from 3D flux maps. The dashed lines are from maps that are generated
from the density maps.}
\label{fig:3Dflux_bias}
\end{figure}

Using the transformation given in Eq.(\ref{eq:transform}), we can express the pdf $p(F_{\alpha})$ and bias $b(< F_{\alpha})$ in terms of the
pdf $p(\delta_0)$ and bias for the density $b(>\delta_0)$:
\be
b(>\delta_0) \equiv {1 \over \sqrt {\xi^{\delta}_{12}}}\left [ {\int_{\delta_0}^{\infty} \int_{\delta_0}^{\infty} p(\tilde\delta_1,\tilde\delta_2) 
d\tilde \delta_1 d\tilde \delta_2 \over 
[\int_{\delta_0}^{\infty} p(\tilde\delta) d\tilde\delta ]^2} -1 \right ]^{1/2} =
{1 \over \sqrt {\xi^{\delta}_{12}}}\left [ {\int_{0}^{F_{\alpha}} \int_{0}^{F_{\alpha}} p(F^{(1)}_{\alpha},F^{(2)}_{\alpha}) 
dF^{(1)}_{\alpha} dF^{(2)}_\alpha \over 
[\int_{0}^{F_{\alpha}} p(F_{\alpha}) dF_{\alpha} ]^2} -1 \right ]^{1/2} 
\equiv \sqrt{\xi_\alpha \over \xi_\delta } \;\; b(< F_{\alpha})
\label{eq:flux_bias}
\ee
The results of such calculation for the PDF is shown in Figure (\ref{fig:3Dflux_pdf}).
There is no accurate model for bias in the highly nonlinear regime. The lognormal approximation is accurate in the quasilinear regime
and starts to fail in the highly nonlinear regime. The approach taken by \citep{VaMu04} can be extended to the construction of bias.
The results of integrated bias for the flux as defined in  Eq.(\ref{eq:flux_bias}) and are plotted in Figure (\ref{fig:3Dflux_bias}). 
We found that including a power law mapping $(1+\delta)\rightarrow (1+\delta)^{\Delta}$, where $\Delta$ is a constant of order unity, can fit the numerical results, if it is applied before 
the exponential transformation. 
\section{Projected Flux Decrement}
We are interested in scales larger than the comoving Jeans length. At around redshift $z=3$ which is about $1h^{-1} {\rm Mpc}$.
The fractional flux  ${\myF}_{\alpha}(z,\oh)$ in this limit can be related to the underlying dark matter density distribution $\delta$,
using Eq.(\ref{eq:flux}) by the following expression:
\be
\delta{F}_{\alpha}(z,\oh) \equiv (F_{\alpha}(z,\oh)-\langle F_{\alpha}(z,\oh) \rangle)/\langle F_{\alpha}(z,\oh) \rangle = -A(z)\beta(z) \tilde\delta(\oh,z) \approx -A(z) \beta(z)b(z) \delta(\oh,z).
\label{eq:fit}
\ee
The relation assumes that IGM traces the dark matter distribution at larger scales (above Jeans length). 
We have incorporated an additional redshift dependent bias factor $b(z)$ in relating $\delta_{\rm IGM}(z)$ and $\delta(z)$
for generality. The results presented here can be generalised for arbitrary redshift dependent bias. In our calculations we will set $b(z)=1$.  
The factors $A(z)$ and $\beta$ have already been introduced above;
see e.g. \cite{HG97} and \cite{Mc03} for a discussion on the temperature-density relation which is well described
by a power law. 

The integrated flux decrement $\myF_{\alpha}(\oh)$ towards a particular direction $\oh$ can be expressed 
through a line-of-sight integration as a projected density contrast:
\be
\delta{\cal F}_{\alpha}(\oh) = 
\int_{r_i}^{r_q} dr\;\delta F_{\alpha}(\oh,r) \approx \int_{r_i}^{r_q} dr A(r)\beta(r) \tilde\delta_{}(\oh,r) 
\equiv \int_{r_i}^{r_q} \; dr \; \omega_{\alpha}(r) \; \tilde\delta_{}(\oh,r). 
\ee
The range of comoving distances probed by the Lyman-$\alpha$ spectrum extends from $r_i$ to the quasar distance $r_q$. 
We will define a parameter $\myF_{\alpha}^{\rm min} = -\int_{r_i}^{r_q}\;w_{\alpha}(r)\; dr$ that is the minimum value of $\myF_{\alpha}$ for a given $r_i$ and $r_q$. The weight $w_{\alpha}(z)$ that acts as a kernel
for projection can be expressed in terms of previously introduced functions $w(z)=A(z)\beta(z)$.
Few comments are in order. The assumption of quasilinear theory (e.g. the lognormal model) is expected to be valid above the Jeans length.
This is where our results are expected to be valid \citep{EJ98}. The relationship between the flux and the 
underlying mass distribution has been verified in numerous studies \citep{MLZ09,BD97,Croft02, Viel02, Saitta08}. 
It is also important to note that the simple linear relationship that is often
used in deriving many analytical results in the context of Lyman-{$\alpha$} studies 
can be modified in the presence of any non-gravitational process such as fluctuations
in the level of ultraviolet radiation or temperature fluctuation. Such non-gravitational
effects are harder to model analytically. Next we will use these expressions to model 
the lower order statistics of $\delta{\cal F}_{\alpha}(\oh)$.
\subsection{Lower order Cumulants and Cumulant Correlators in Projection}
\label{sec:lower}
We will use both cumulants and cumulants correlators in our studies. The one-point cumulants
can be estimated from a single LOS where as neighbouring pairs of QSOs are required
to probe the cumulant correlators. By construction, the normalised cumulant correlators are independent
of the separation of quasar pairs. Increasing the number of line sights will help to increase the
signal-to-noise.

To compute the variance and other lower order moments or cumulants
we start with the Fourier transform of the 3D density contrast $\delta$ which we denote as $\delta(\bk)$:
\be
\myF_{\alpha}(\oh) = \inc {dr} \; \omega_{\alpha}(r) \int {d^3{\bf k} \over {(2
\pi)}^3} \exp ( i r k_{\parallel} + i d_A(r){{\bf \theta}_{12}}\cdot {\bf k}_{\perp} ) \delta(\bk,r).
\ee
Here ${\bf \theta}_{12}$ denotes the angle between
the line of sight direction $\oh$ and should be treated as a vector on the surface of the sky and the wave vector ${\bf
k}$,  $k_{\parallel}$ and ${\bf k}_{\perp}$ denote the components
of ${\bf k}$, parallel and perpendicular to the line of sight direction.
In the small angle approximation one assumes that $|{\bf k}_{\perp}|
>> k_{\parallel}$.
Using these definitions we can compute the projected variance $\langle \myF^2_{\alpha}\rangle$
in terms of the dark matter power spectrum
$P_{\delta}(k,r)$ \citep{Kaiser92}:
\begin{equation}
\langle \myF_{\alpha}^2(\oh) \rangle_c = \inc d {r}
{\omega_{\alpha}^2(r) \over d_A^2(r)} \int {d^2 {\bf l} \over (2
\pi)^2}~ {\rm P}_{\delta}{ \Big ( {l\over d_A(r)}, r \Big )} 
\label{alpha_variance}
\end{equation}
Similarly the higher order moments of the field
relate $\langle \myF^p_{\alpha}\rangle$ to the
three-dimensional multi-spectra of the underlying dark matter
distribution ${\rm B}_p$ (Hui 1999; Munshi \& Coles 1999a):
\be
\langle \myF_{\alpha}^p(\oh) \rangle_c = \inc d {r}
{\omega^p_{\alpha}(r) \over d_A^{2(p-1)}(r)}\int {d^2 {\bf l_1} \over (2\pi)^2} \cdots
\int {d^2 {\bf l_p} \over (2 \pi)^2}{\rm B}_{\delta}^{(p)} \left ( {l_1\over d_A(r)}, \cdots, {l_p\over d_A(r)} \right )_{\sum {\bl}_i=0}.
\ee
The subscript $c$ denotes the connected part of a diagram that represents the higher order correlation function;
Limbers approximation \citep{Limb54} is used to derived the above result. The subscript $(\sum {\bl}_i=0)$ denotes
a multiplicative Dirac's delta function $\delta_{\rm D}[\sum_{i=1}^p({\bl}_i)]$.
We have used the compact notation ${\rm B}_{\delta}^{(p)}$ to denote 
the $p$-th order multispectra where  ${\rm B}_{\delta}^{(3)}$ denotes
the bispectrum and  ${\rm B}_{\delta}^{(4)}$ denotes the trispectrum, often
denoted as ${\rm T}_{\delta}$ in the literature. The power spectrum correspond to ${\rm B}_{\delta}^{(2)}$.
We will use these results to show that it is possible
to compute the complete probability distribution function (PDF)
of $\myF_{\alpha}$ from the underlying dark matter PDF. 
The details of the analytical results presented
here can be found in \citep{MuJa00,MuJa01}.

In addition to the cumulants their correlators are important as 
they are related to the two-point PDF (or 2PDF) and hence the bias
associated with the higher flux regions
\be
\langle \myF_{\alpha}^p(\oh_1)\myF_{\alpha}^q(\oh_1) \rangle_c = \inc d {r}
{\omega^{p+q}_{\alpha}(r) \over d_A^{2(p+q-1)}(r)}\int {d^2 {\bf l_1} \over (2\pi)^2} \cdots
\int {d^2 {\bf l}_{p+q} \over (2 \pi)^2}{\rm B}_{\delta}^{(p+q)} \left ( {l_1\over d_A(r)}, \cdots,{l_{p+q}\over d_A(r)}  \right )_{\sum_i {\bl_i}=0}.
\ee
A special case of this equation corresponds to the expression for the two-point correlation 
function $\langle \myF_{\alpha}^p(\oh_1)\myF_{\alpha}^q(\oh_1) \rangle$. T
are generic. We will specialise them by assuming a particular form for the 
higher order multispectra to make further progress. Just as the normalised cumulants $S_p$
we can define the normalised cumulant correlators 
$C_{pq}=\langle\myF_{\alpha}^p(\oh_1)\myF_{\alpha}^q(\oh_1)\rangle/
\la\myF_{\alpha}(\oh_1)\myF_{\alpha}(\oh_1)\rangle\ra\la\myF^2_{\alpha}(\oh)\ra$.
We will use the normalised cumulants or the $S_p$ parameters and the normalised cumulant correlators to
construct the PDF and the bias associated with the flux decrements or $\myF_{\alpha}$. We will
denote the normalised cumulants and their correlators of density by $S_p$ and $C_{pq}$.
The corresponding quantities for the flux-decrements will be denotes using a superscript $\alpha$ i.e $S_p^{\alpha}$ and $C_{pq}^{\alpha}$.
\subsection{A Minimal Tree-Model}
\label{sec:hier}
The spatial length scales, corresponding to the small angular scales,
that are relevant in our discussion, are in the highly non-linear regime. Assuming a
\emph{tree model }for the matter correlation hierarchy in the highly
non-linear regime, one can write the general form of the $N$th order
correlation function $\xi^{(p)}_{\delta}$ as a product of two-point
correlation function $\xi^{(2)}_{\delta}$ \citep{GP77, DP77, Fry84, BS89, szsz93,szsz97}
In Fourier space such an \emph{ansatz} means that the hierarchy of multispectra can be
written as sums of products of the matter power-spectrum:
\begin{eqnarray}
&&B_\delta({\bf k}_1, {\bf k}_2, {\bf k}_3)_{\sum k_i = 0} = Q_3( P_{\delta}({k_1})P_{\delta}({k_2}) + P_{\delta}({k_2})P_{\delta}({k_3})
+ P_{\delta}({k_3})P_{\delta}({k_1}) ) \label{eq:hier1} \\
&&T_\delta({\bf k}_1, {\bf k}_2, {\bf k}_3, {\bf k}_4)_{\sum k_i = 0} = R_a
\ P_{\delta}({k_1})P_{\delta}(|{\bf k_1 +
k_2}|) P_{\delta}(|{\bf k_1 + k_2 + k_3}|)  + {\rm cyc. perm.} + R_b \ P_{\delta}({k_1})P_{\delta}({k_2})P_{\delta}({k_3}) +
{\rm cyc. perm.} 
\label{eq:hier2}
\end{eqnarray}
The subscript $({\sum k_i = 0})$ represents a Dirac's $\delta$ function $\delta_D[\sum_{i=1}^p(\bk_i)]$. 
Different hierarchical models differ in the way they specifiy the amplitudes $Q_3,R_a,R_b$ etc. The results
that we will derive are valid for generic hierarchical models.
Using Eq.(\ref{eq:hier1}) and Eq.(\ref{eq:hier2}) we can write:
\begin{equation}
\langle \myF_{\alpha}^3\rangle_c = (3Q_3){\cal C}_3[(\myJ^{\alpha}_0(r))^2]
= S_3^{\alpha} \langle \myF_{\alpha}^2 \rangle_c^2 \quad\quad
\langle \myF_{\alpha}^4\rangle_c = (12R_a + 4
R_b){\cal C}_4[(\myJ_0^{\alpha}(r))^3] = S_4^{\alpha} \langle \myF_{\alpha}^2 \rangle^3_c,
\label{eq:s3s4}
\end{equation}
Where we have defined the following quantities:
\begin{equation}
{\cal C}_p\left [(\myJ^{\alpha}_0(r))^{p-1}\right ] = \int_{r_i}^{r_q} {\omega_{\alpha}^p(r)
\over d_A^{2(p-1)}(r)}(\myJ^{\alpha}_0(r))^{(p-1)} dr; \quad\quad
[\myJ^{\alpha}_0(r)] =  \int  \frac{d^2\bf l}{(2\pi)^2} P_
{\delta}
\left( {l \over d_A(r)} \right).
\end{equation}

These lower order moments are next used to construct the {\em void} probability function (VPF)
which also acts as a generating function for the normalised $S_p^{\alpha}$ parameters for flux-decrement $\myF_{\alpha}(\oh)$.
The probability distribution function (PDF) can be constructed from the knowledge of the VPF as will
be detailed in the next section.
\begin{figure}
\begin{center}
{\epsfxsize=6. cm \epsfysize=6 cm {\epsfbox[31 416 312 712]{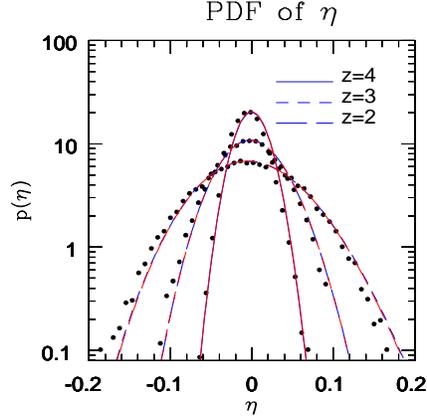 }}}
\end{center}
\caption{The PDF of the projected flux-decrement $p(\myF_{\alpha})$ is plotted. The two different approximations
that we have studied produces near identical results. Three different redshifts are shown. The lower (higher) redshift
corresponds to broader (narrower) distribution. The lines of sight are drawn from a $512^3$ simulation box. A total of $512^2$
LOS are analysed which are distributed on a regular grid.}
\label{fig:2D_pdf}
\end{figure}
\subsection{Probability Distribution Function of Transmitted Flux}
\label{sec:pdf}
Before we discuss our techniques to construct the PDF for the projected transmitted flux few 
explanations regarding the notation and its link to the entire hierarchical paradigm are in order.
Although our primary aim is to construct the PDF it is usually easier to construct the 
moment generating function $\phi(y)$
that will be introduced later. The lower order normalised cumulants (also known as the $S_p$ parameters) up to
an arbitrary order can be recovered, if necessary, using a Taylor expansion of $\phi(y)$. The function $h(x)$
is related to $\phi(x)$ through an inverse Laplace transform, i.e., a knowledge of $S_p$
parameters to an arbitrary order can be used to construct the function $h(x)$. The function $h(x)$ on the other hand 
is simply a scaled PDF and the real PDF can be extracted from it using the definition of scaling variable $x$ 
(to be defined later). 

The construction of PDF for $\myF_{\alpha}$ involves additional steps. First the generating function $\Phi^{\alpha}(y)$ for 
the lower order normalised cumulants $S^{\alpha}_p$ is expressed in terms of the underlying $\phi^{}(y)$. 
Next we define a new scaling variable for $\myF_{\alpha}$ which is denoted by $\eta$. We show that, under certain simplifying
approximation, the cumulant generating function for $\eta$ i.e. $\Phi^{\eta}(y)$ is same as the underlying
cumulant generating function $\phi^{}(y)$. Once this proved, it is easy see that the scaled PDFs  $H^{\eta}(y)$
and $h^{}(y)$ too are same as they are simply the Laplace transform of $\Phi^{\eta}(y)$ and $\phi^{}(y)$
respectively. This technique we use has already been used to derive similar relations in the context of weak lensing
statistics \citep{MuJa00,MuJa01,V00,BV00}.

The generating function $\Phi^{\alpha}(y)$ (or the Void Probability Function, i.e. VPF) for the lower order normalised cumulants $S^{\alpha}_p$ for the flux-decrement $\myF_{\alpha}(\oh)$ 
can be expressed as:
\begin{equation}
\Phi^{\alpha}(y)\equiv \sum_{p=1} {S^{\alpha}_p \over p!}y^p = y+\sum_{p=2}^ {\infty} {{\langle
\myF_{\alpha}^p(\oh) \rangle_c} \over \langle \myF_{\alpha}^2 (\oh)
\rangle_c^{p-1}} y^p.
\end{equation}
Our aim is to relate the VPF $\Phi^{\alpha}(y)$ and $\phi^{}(y)$ where $\phi^{\alpha}(y)$ is the VPF
of the underlying matter distribution. The exact functional form of $\phi(y)$ is fixed by choosing a 
specific hierarchical ansatz. We will keep our discussion here completely general without any reference to any specific form
of the correlation hierarchy. Using Eq.(\ref{eq:s3s4}) we can write:
\begin{equation}
\Phi^{\alpha}(y) = \int^{r_q}_{r_i} \sum_{p=1}^{\infty}
{ 1 \over p!} S_p^{} {\omega_{\alpha}^p(r) \over d_A(r)^{2(p-1)}}
[\myJ^{\alpha}_0(r)]^{ (p-1)} { y^p \over \av_c^{(p-1)} } - y\myF^{\rm min}_{\alpha}; 
\end{equation}
Using the VPF of the underlying mass distribution $\phi(x)$, the above expression can be
rewritten in a more compact form:
\begin{equation}
\Phi^{\alpha}(y) =  \int^{r_q}_{r_i} \;dr\; d_A^2(r)
\Big[ {\langle \myF^2_{\alpha}\rangle_c \over \myJ^{\alpha}_0(r)} \Big
] \phi \Big [ {\omega_{\alpha} (r) \over d_A^2 (r)} {\myJ^{\alpha}_0(r) \over \av_c} y \Big ]; \quad\quad 
\phi(y)=\sum_p{S_p \over p!}y^p
\end{equation}
We will define a variable $\eta$ that will make the analysis simpler. The VPF for $\eta$,  denoted 
$\Phi^{\eta}(y)$, can now be expressed in terms of the VPF of the underlying mass distribution   $\phi(y)$
by the following expression:
\be
\Phi^{\eta}(y) = {1 \over |\myF_{\alpha}^{\rm min}|}\int^{r_q}_{r_i} \;dr\; 
\left[ {d_A^2(r) \langle \myF^2_{\alpha}\rangle_c \over \myJ^{\alpha}_0(r)|\myF_{\alpha}^{min}|} \right
] \phi \left [ |\myF_{\alpha}^{\rm min}|{\omega_{\alpha} (r) \over d_A^2 (r)} {\myJ^{\alpha}_0(r) \over \av_c} y \right ]; \quad\quad
\eta = {(\myF_{\alpha} - \myF_{\alpha}^{\rm min}) / \myF_{\alpha}^{\rm min} }. 
\ee
The scaling properties of the PDF in an hierarchical model are encoded in a scaling function. The scaling
function $H^{\eta}(x)$ is related to the PDF of $\eta$ i.e.$ p(\eta)$ where as the scaling function $h(x)$
is associated with the PDF of the underlying mass distribution $\delta$. The following relation 
relates $H^{\eta}(x)$ and the scaling function $h(x)$.
\be
H^{\eta}(x) = {1 \over |\myF_{\alpha}^{\rm min}| }\int^{r_q}_{r_i}\; w_{\alpha}(r) dr\; 
\left [{d_A^2(r) \la \myF^2_{\alpha}\ra_c \over \myJ^{\alpha}_0(r) | \myF_{\alpha}^{\rm min}| w_{\alpha}(r)} \right ]^2
h\left [x {\la \myF^2_{\alpha} \ra_cd_A^2(r)}\over w_\alpha(r)|\myF^{\rm min}_{\alpha}|{\cal J}^{\alpha}_0(r) \right ].
\ee
The above relation is derived using the following definition of $H^{\eta}(x)$ in terms of the moment generating
function or VPF $\Phi^{\eta}(y)$ \citep{BS89} as the moment generating function $\Phi^{\eta}(y)$ and the scaled PDF $H^{\eta}(x)$
are linked through an inverse Laplace transform:
\be
H^{\eta}(x) = -\int^{i\infty}_{-i\infty} {dy \over 2\pi i} \exp(xy) \Phi^{\eta}(y) 
\ee
A similar result holds for $h(x)$ and $\phi(y)$. Using an approximate form for the integrals we recover the following relation:
\be
p_{\alpha}(\myF_{\alpha}) = {p(\eta)/|\myF^{\rm min}_{\alpha}|}
\label{eq:pdf_final}
\ee
In terms of the cumulants
this means $S^{\alpha}_p = S_p/|\myF^{\rm min}_{\alpha}|^{p-2}$. 
It is instructive to note that the final result is independent of any detailed modelling. We will use a lognormal
PDF as well as the PDF constructed using the hierarchical ansatz to model  $p(\eta)$. 

It is important to note that the function $h(x)$ is fundamental to all scaling analysis. It was initially introduced
to study the clustering of galaxies later was used in diverse cosmological studies (e.g. \cite{VSS99}); e.g. the
mass function of collapsed objects that includes clusters, groups and galaxies can all be described the
scaling function $h(x)$. The multiplicity function $\eta(M,z)$ at a redshift $x$ for collapsed objects 
of mass $M$ can be expressed via: $\eta(M,z)dM/M = (\bar\rho/M) x h(x) dx$. Here the scaling
function $x$ can be expressed as $x=[1+\delta(M,z)]/\bar \xi_2$ and $\bar\rho$ is the mean physical 
density of the Universe. Thus, scaling arguments can also provide an alternative to the usual mass function
calculation based on the Press-Schecter formalism. These issues have been discussed extensively in the literature 
(e.g. \cite{VS00}).
The volume averaged two-point correlation function $\bar\xi_2$ is evaluated at a corresponding relevant scale.
It is interesting to notice that the functional form for the scaling function $h(x)$ depends only
on the form for initial power spectrum and its qualitative behaviour can be predicted using very general arguments.
Two different asymptotes are particularly well understood;   (1) $x\ll1 : h(x) \propto x^{\omega-2}$ and (2) $x\gg1 : h(x) \propto x^{\omega_s-1}
\exp(-x/x_{\star})$ with $\omega \sim 0.5$ and $\omega_s \sim -3/2$. More detailed discussions about
some of the salient features are reported in Appendix-A.

The numerical results for the 2D PDF of $\eta$ are presented in Figure(\ref{fig:2D_pdf}). The results are depicted 
for three redshifts, $z=2,3,4$. These results were computed using a grid of $512^2$ LOS drawn from a 
$512^3$ simulation box.The underlying cosmology is same as the 
ones adopted for 3D studies. The resulting PDFs computed using this grid are stable down to $O(10^{-4})$.
Notice that the projected PDF is closer to a Gaussian in form than its 3D counterpart. 
\subsection{Joint PDF of Flux-Decrements of neighbouring Quasar Spectra}
The  2PDF is constructed from the cumulant correlators. The cumulant correlators
can be constructed by correlating flux-decrements of neighbouring lines of sight $\oh_1$ and  $\oh_2$. The 
cumulant correlators are two-point statistics and generalises the usual two-point correlation function to
higher order that have the ability to probe non-Gaussianity. More interestingly they are related 
to the bias associated with high flux-decrement regions.
\begin{equation}
C^{\alpha}_{pq} =
{\langle \myF_{\alpha}(\oh_1)^p \myF_{\alpha}(\oh_2)^q \rangle_c/\langle  \myF_{\alpha}^2(\oh) \rangle_c^{p+q-2} \langle
\myF_{\alpha}(\oh_1)\myF_{\alpha}(\oh_2) \rangle_c }.
\end{equation}
Notice that, by construction, $C_{11}=1$ which is due to the particular normalization that is generally adopted for $C_{pq}$. 
At the lower order the cumulant correlators can be expressed in terms of the hierarchical amplitudes (i.e. $Q_3, R_a,R_b,S_a,S_b,S_c$): 
\begin{eqnarray}
&& \langle {\delta\cal F}_{\alpha}^2(\oh_1){\delta\cal F}_{\alpha}(\oh_2) \rangle_c  = 
2Q_3 {\cal C}_3 [\myJ^{\alpha}_{0} \myJ^{\alpha}_{12}] =
C_{21}^{\eta}{\cal C}_3 [\myJ^{\alpha}_{0} \myJ^{\alpha}_{12}] \equiv C_{21}^{\alpha} \langle
{\myF}_{\alpha}^2 \rangle_c \langle \delta{\cal F}_{\alpha}(\oh_1)\delta{\cal F}_{\alpha}(\oh_2) \rangle_c, \\
&& \langle{\delta\cal F}_{\alpha}^3(\oh_1) {\delta\cal F}( \oh_2) \rangle_c  = 
(3R_a + 6 R_b){\cal C}_4 [(\myJ^{\alpha}_{0})^2 \myJ^{\alpha}_{12}] =
 C_{31}^{\eta}{\cal C}_4 [(\myJ^{\alpha}_{0})^2 \myJ^{\alpha}_{12}]
 \equiv  C_{31}^{\alpha} \langle
{\delta\cal F}_{\alpha}^2 \rangle_c^2 \langle {\cal F}(\oh_1) {\cal F}(\oh_2) \rangle_c
,\\ &&  \langle {\cal F}_{\alpha}^2(\oh_1) {\cal F}_{\alpha}^2(\oh_2) \rangle_c  =
4 R_b{\cal C}_4 [(\myJ^{\alpha}_{0})^2 \myJ^{\alpha}_{12}]
= C_{22}^{\eta}{\cal C}_4 [(\myJ^{\alpha}_{0})^2 \myJ^{\alpha}_{12}]
\equiv  C_{22}^{\alpha} \langle
{\cal F}^2 \rangle_c^2 \langle {\delta\cal F}_{\alpha}(\oh_1){\delta\cal F}_{\alpha}(\oh_2) \rangle_c ,\\
&& \langle {\delta\cal F}_{\alpha}^4(\oh_1) {\delta\cal F}_{\alpha}(\oh_2)\rangle_c  = 
(24S_a + 36S_b + 4 S_c){\cal C}_5 [(\myJ^{\alpha}_{0})^3 \myJ^{\alpha}_{12} ] =
C_{41}^{\eta} {\cal C}_5 [\myJ^{\alpha}_{0} \myJ^{\alpha}_{12}]
\equiv  C_{41}^{\alpha} \langle
{\delta\cal F}_{\alpha}^2 \rangle_c^3 \langle {\delta\cal F}_{\alpha}(\oh_1){\delta\cal F}_{\alpha}(\oh_2)
\rangle_c
,\\ && \langle {\delta\cal F}_{\alpha}^3(\oh_1) {\delta\cal F}_{\alpha}^2(\oh_2) \rangle_c = 
 (12S_a + 6 S_b){\cal C}_5[(\myJ^{\alpha}_{0})^3 \myJ^{\alpha}_{12}] =
C_{32}^{\eta}{\cal C}_5[(\myJ^{\alpha}_{0})^3 \myJ^{\alpha}_{12}]
\equiv  C_{32}^{\alpha} \langle
{\delta\cal F}_{\alpha}^2 \rangle_c^3 \langle {\delta\cal F}_{\alpha}(\oh_1){\delta\cal F}_{\alpha}(\oh_2)
\rangle_c.
\end{eqnarray}
The expressions for the cumulant correlators given above apply in the large separation limit, i.e. $\la \myF_{\alpha}({\oh_1})\myF_{\alpha}({\oh_2}) 
\ra\ll\la\myF_{\alpha}^(\oh)\ra$.
Though higher-order correction terms can be computed using the framework of the hierarchical ansatz, in practice 
the large separation limit is reached very quickly even for nearby lines of sight.
A detailed discussion of the tree-amplitudes can be found in \citep{MCM3}. In general the cmulant correlators can be expressed as:
\be
\la {\delta\cal F}_{\alpha}^p(\oh_1) {\delta\cal F}_{\alpha}^q(\oh_2) \ra_c
\equiv C_{pq}^{\eta}{\cal C}_{p+q}[(\myJ_0^{\alpha}(r))^{p+q-2} \myJ^{\alpha}_{12}(r)]
= C_{pq}^{\alpha} \langle{\delta\cal F}_{\alpha}^2 \rangle_c^{p+q-2} \langle {\delta\cal F}_{\alpha}(\oh_1){\delta\cal F}_{\alpha}(\oh_2)\rangle_c.
\label{eq:cumu}
\ee
\begin{figure}
\begin{center}
{\epsfxsize=6 cm \epsfysize=6 cm {\epsfbox[312 416 591 712]{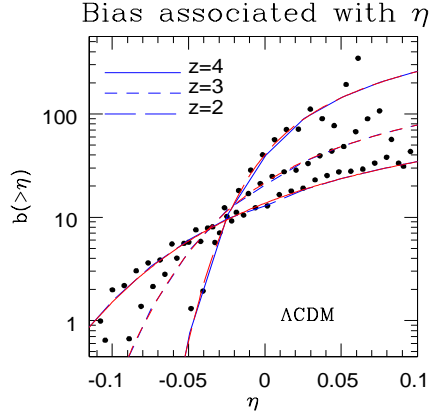 }}}
\end{center}
\caption{The cumulative bias $b(>\myF_{\alpha})$ of the flux is plotted as a function of the flux. The log-normal distribution
and the hierarchical ansatz generates near identical results. For a given positive $\delta F_{\alpha}$ the bias is typically higher for a 
higher redshift. The bias $b(>\myF_{\alpha})$  samples the underlying 3D bias associated with $\tilde\delta$ i.e. b($\tilde\delta$).}
\label{fig:sim}
\end{figure}
\n
Using these expressions we can construct the joint two-point PDF or the bias. To do so we introduce the generating function 
$\beta^{(2)}_{\alpha}(y_1, y_2)$ for the normalised cumulant correlators:
\begin{equation}
\beta^{(2)}_{\alpha}(y_1, y_2) =\sum_{p,q}^{\infty} {y_1^py_2^q \over p! q!}C^{\alpha}_{pq} = \sum_{p,q}^{\infty} {1 \over p! q!} { y_1^p
y_2^q\over
\langle \myF_{\alpha}^2\rangle_c^{p+q-2} } {\langle\myF_{\alpha}(\oh_1)^p\myF_{\alpha}(\oh_2)^q \rangle_c \over \langle
\myF_{\alpha}(\oh_1)\myF_{\alpha}(\oh_2)\rangle_c }.
\end{equation}
Using the expression Eq.(\ref{eq:cumu}) derived using the hierarchical ansatz we can make further progress:
\be
\beta^{(2)}_{\alpha}(y_1, y_2) = \sum_{p,q}^{\infty} {C^{\alpha}_{pq} \over p! q! } { y_1^p \over
\langle \myF_{\alpha}^2\rangle_c^{p-1}}{ y_2^q \over \langle \myF_{\alpha}^2 \rangle_c^{q-1} } { 1 \over \xi_{\alpha}^{12}}  \int_0^{r_s}\; dr\; d_A^2(r){\omega_{\alpha}^{p}(r) \omega_{\alpha}^{q}(r) \over d_A(r)^{2p} d_A(r)^{2q} } [{\myJ}^{\alpha}_{0}(r)]^{p+q-1}{\myJ}^{\alpha}_{{12}}(r).
\ee
Next we introduce the generating function $\beta^{(1)}_{\alpha}(y)$ for the normalised cumulant correlators. While 
$\beta^{(2)}_{\alpha}(y_1, y_2)$ is associated with $C_{pq}$ the generating function $\beta^{(1)}_{\alpha}(y)$ is the generating function for $C_{p1}$
In 3D the hierarchical ansatz ensures the fcatorization property $C_{pq}=C_{p1}C_{q1}$. 
\be
\beta^{(2)}_{\alpha}(y_1, y_2) = \inc \; dr \; {w^2_{\alpha}(r)}  \; d^2_A(r)\;
{ \myJ^{\alpha}_{{12}} \over \xi_{\alpha}^{12}}
{ \la \myF_{\alpha}^2\ra_c \over {\myJ^{\alpha}_{0}(r)}}
\beta_{\alpha} \Big ({ y_1 \over \la \myF_{\alpha}^2 \ra_c} {\omega_{\alpha}(r) \over d^2_A(r)}{\myJ}^{\alpha}_0(r) \Big )
{ \la \myF_{\alpha}^2\ra_c
\over {{\myJ}^{\alpha}_0}(r)} \beta^{(1)}_{\alpha} \Big ({ y_2 \over \la \myF_{\alpha}^2 \ra_c} {\omega_{\alpha}(r) \over d_A^2
(r)} {\myJ}^{\alpha}_0(r)  \Big ); \quad \beta^{(1)}_{\alpha}(y) = \sum_{p=1}^{\infty} {C^{\alpha}_{p1} \over p!}y^p.
\ee
In terms of the reduced flux $\eta$ the generating function for the joint cumulant correlator can be written as:
\be
\beta^{(2)}_{\eta}(y_1, y_2) = \inc \; dr \;{ d^2_A(r)\over |\myF_{\alpha}^{\rm min}|^2 }\;
{ \myJ^{\alpha}_{{12}}(r) \over \xi_{\alpha}^{12}}
{ \la \myF_{\alpha}^2\ra_c \over {\myJ^{\alpha}_{0}}(r)}
\beta^{(1)}_{\eta} \Big ({ y_1 \over \la \myF_{\alpha}^2 \ra_c} {\omega_{\alpha}(r) \over d^2_A(r)}{\myJ}^{\alpha}_0(r) |\myF_{\alpha}^{\rm min}| \Big )
{ \la \myF_{\alpha}^2\ra_c
\over {{\myJ}^{\alpha}_0}(r)} \beta^{(1)}_{\alpha} \Big ({ y_2 \over \la \myF_{\alpha}^2 \ra_c} {\omega_{\alpha}(r) \over d_A^2
(r)} {\myJ}_{0}^{\alpha}(r) |\myF_{\alpha}^{\rm min}| \Big ).
\ee
These equations are generic and exact. They only depend on the hierarchical ansatz for the higher order correlation
functions. To simplify further we replace the relevant integrals by the following approximate results:   
\begin{eqnarray}
 \bar\xi_{\alpha}  \approx {1\over 2} \Delta r  {\omega_{\alpha}^2(r_c)\over d_A^2(r_c)} \Big [ \int {d^2 l \over (2\pi)^2} 
{{\rm P}_{\delta}({l \over d_A(r_c)})} \Big ]; \quad\quad
\xi_{\alpha}^{12} \approx {1\over 2} \Delta r  {\omega_{\alpha}^2(r_c) \over d^2_A(r_c)} \Big [ \int {d^2 l \over (2\pi)^2} 
{{\rm P}_{\delta}({l \over d_A(r)})} \exp [i {\bf \;l} \cdot \theta_{12}] \Big ]; \quad\quad \Delta r=r_q-r_i.
\end{eqnarray}
This leads to the following expressions which also preserves the factorization properties for the generating functions:
\be
\beta^{(2)}_{\eta}(y_1,y_2) = \beta^{(1)}_{\eta}(y_1)\beta^{(1)}_{\eta}(y_2).
\ee
From the generating function $\beta^{(1)}_{\eta}(y)$ it is possible to construct the scaling function $b_{\eta}(x)$ that
encodes all the information about the bias $b(\eta)$. The functions $b_{\eta}(x)$ can be recovered using the following
integrals in the complex $y$ plane \citep{BS89}: 
\begin{equation}
b_{\eta}(x) h_{\eta}(x) = -{ 1 \over 2 \pi i}
\int_{-i\infty}^{i\infty} dy \beta^{(1)}_{\eta}(y) \exp (xy); \quad\quad
b_{\eta}(>x) h_{\eta}(>x) = -{ 1 \over 2 \pi i}
\int_{-i\infty}^{i\infty} dy {\beta^{(1)}_{\eta}(y)\over y} \exp (xy).
\end{equation}
In addition to the usual scaling functions $h(x)$ and $b(x)$ that are associated with the scaling properties of PDF $p(\eta)$ and bias $b(\eta)$
we also include their cumulative versions $h(>x)$ and $b(>x)$ which are linked to $p(>\delta)$ and $b(>\delta)$ i.e. the PDF and
bias beyond a threshold $\delta$. Using the hierarchical ansatz allows us to write the 2PDF using the following factorized form:
\be
p(\eta_1,\eta_2)= p(\eta_1)p(\eta_2)(1+ b(\eta_1)\xi_{12}^{\eta}b(\eta_2))d\eta_1d\eta_2.
\ee
Finally, using the definition of $\eta$ we can also write the 2PDF of the flux decrement $\myF_{\alpha}$ for two-different
neighbouring lines of sight:
\be
p_{\alpha}(\myF_{\alpha}(\Omega_1),\myF_{\alpha}(\Omega_2))= p_{\alpha}(\myF_{\alpha}(\Omega_1))p_{\alpha}(\myF_{\alpha}(\Omega_2))(1+ b_{\alpha}(\myF_{\alpha}(\Omega_1))\xi_{12}^{\alpha}b_{\alpha}(\myF_{\alpha}(\Omega_2)))
d\myF_{\alpha}(\Omega_1)d\myF_{\alpha}(\Omega_2)
\ee
The bias $b_{\alpha}$ for flux-decrement and the bias associated with the underlying mass distribution $\eta$ are related
by the following expression:
\be
b_{\alpha}(\myF_{\alpha}) = {b_{\eta}(\eta) / |\myF_{\alpha}^{\rm min}|}.
\label{eq:bias_final}
\ee
This approximate result also means that we can write $C^{\alpha}_{p1} = C^{\eta}_{p1}/|\myF_{\alpha}^{min}|^{p-1}$; which
implies that  $C^{\alpha}_{pq} = C^{\eta}_{pq}/|\myF_{\alpha}^{min}|^{p+q-2}$. 
Together, Eq.(\ref{eq:pdf_final}) and Eq.(\ref{eq:bias_final}) are sufficient to model the one and two-point distributions
of transmitted flux. The fact that the final results are completely independent of specific models for the
PDF of underlying density contrast indicates their general validity;  we can use both the hierarchical model
as well as the lognormal distribution as a model for the underlying density contrast $\delta(\bx) = \eta(\bx)-1$.

Though we have only provided the joint PDF of two lines-of-sight directions in this article the formalism 
can be extended to tackle arbitrary number of line of sights. The statistics derived using multiple lines of sight 
are directly linked to the statistics of {\em hot-spots} in flux-decrement maps.
It is interesting to notice that the bias $b(x)$ introduced here has also been used in previous studies 
that dealt with clustering of Lyman-{$\alpha$} absorbers. In a more general context the correlation function of two over-dense 
objects $\hat\xi_{12}$ can be expressed in terms of the underlying correlation function as:
$\hat\xi_{12}(r_1,r_2;x_1,x_2) = b(x_1)b(x_2)\xi_{12}(r_1,r_2)$. Where scaling paramters $x_1$ and $x_2$ are the scaling
parameters associated with collapsed objects. This is 
a generic outcome of any hierarchical formalism and have been used successfully to compute the bias associated with
collapsed objects from clusters and groups to galactic size halos \citep{VS00}. As was the case for the scaling function
$h(x)$ we quote two asymptotes for the bias function $b(x)$ given by \citep{BeS92}; (1) for $x \ll1;  b(x) \propto x^{(1-\omega)/2}$; 
(2) for $x \gg1$; $b(x) \propto x$.
A more detailed discussion is given in Appendix-A. The bias function $b(x)$ is independent of scale, as was the case for the 
scaled PDF $h(x)$. Information regarding a specific scale can be extracted by evaluating the scaling variable $x$ for that scale.

We note here that lognormal model (see Appendix-\ref{logn}) does not belong to the class of other scale-invariant or scaling models. However 
the joint PDF can also be accurately computed using the lognormal model, and in the large separation limit the lognormal model
to can be employed to compute the integrated bias.

The computation of the 2PDF or the equivalently the bias was done using a $512^2$ grid. We computed the number of LOSs
where the Flux decrement crosses a given threshold which gives us the quantity $\int_{\delta_0}^{\infty} \int_{\delta_0}^{\infty} p(\delta_1,\delta_2)
d\delta_1d\delta_2$. We also use the fractions of LOSs where $\myF_{\alpha}$ crosses a threshold to compute $\int_{\delta_0}^{\infty} p(\delta)d\delta$.
Next using the expression given in Eq.(\ref{eq:int_bias}) we compute the resulting bias which is presented in Figure (\ref{fig:sim}).
\section{Conclusions}
\label{sec:conclu}
The diffuse intergalactic medium acts as a significant reservoir of baryons at low to intermediate redshift 
($z<5$), which can be probed via the absorption lines in the spectrum of distant QSOs. The Lyman-$\alpha$ absorption lines along the line of sights
are due to a vast range of completely different classes of object. These objects include under-dense neutral 
hydrogen clouds, halos of large and over-dense systems which may have already reached virial equilibrium,
as well as UV heated systems that are strongly coupled to their environments. Previous studies have mainly 
dealt with the problem of detailed modelling
of number of these objects as a function of their clustering and internal properties.
Using self-consistent scaling models, which have a long history and were initially employed in galaxy clustering
statistics, several authors have computed the column density distribution of Lyman-$\alpha$ absorption systems
not only for the low column density Lyman-$\alpha$ forest systems but also for Lyman-limit systems and
the damped absorption systems.  In addition to the hierarchical modelling,
lognormal approximation is also quite successful in reproducing the clustering statistics 
of Lyman-{$\alpha$} absorption lines. In this study we showed the approach taken by the lognormal
approximation and the hierarchical ansatz generates near identical results. While previous
studies focussed only on one-point statistics we extend these results to the two-point
distributions and their lower order moments.

In addition to the statistics of absorption lines, the statistics of the transmitted flux play
an important complementary role in Lyman-$\alpha$ studies. The flux can be treated as a continuous field.
Various statistics which are often employed in analysing
the flux include the LOS power spectrum measurements, estimation of bispectrum and more 
recently the entire PDF. The flux PDF contains the information regarding cumulants to an arbitrary order.
Clearly the flux statistics and the column density distributions are related. 
One major goal in this study was to unify these two pictures in the context of the hierarchical ansatz or scaling
models as well as the lognormal approximation. 

The hierarchical model is primarily valid at smaller scales where the correlation functions assumes
a hierarchical form. Gravitational clustering is known to develop such a form of hierarchy both in the 
perturbative (quasi-linear) as well as in the highly non-linear regime. The hierarchical ansatz can be used to describe
the mass functions and the bias associated with collapsed objects using the scaling function $h(x)$ and $b(x)$; the variable
$x$ is a scaling variable. Previous studies that have employed the hierarchical ansatz have shown that the column density 
distribution of absorption systems can be described using the same functional form for $h(x)$ and $b(x)$
that are often used for wide ranging studies from the galaxy clustering to thermal Sunyaev-Zel'dovich effects or
X luminosity of clusters of galaxies. We showed that the same analytical framework can also be
used to understand the statistics of QSO flux measurements, thereby providing a unified statistical approach to
what at first sight appear to be very different observables.

We have approached the modelling of the flux PDF in two different ways. For the 3D analysis,
the PDF of the neutral hydrogen density contrast $\hat\delta$ is modelled according to the hierarchical
ansatz as well as lognormal distribution. We find very good agreement with the numerical simulations for
the entire range of redshifts that we have considered. The predictions from both these models are almost identical
and differ only marginally in the less interesting under-dense regions. We also employ a modified 
version of scale invariant hierarchical approximation which was developed recently \citep{VaMu04}. This
particular approximation provides a very accurate model for the clustering of $\hat\delta$.
Using the fluctuating Gunn-Peterson approximation we map the $\hat\delta$ PDF to that of the 
transmitted flux $F_{\alpha}$. We find reasonable approximation for the allowed range of parameters
$A(z)$ and $\beta$ that define the Gunn-Peterson approximation. Next we consider the projected or the 2D
distribution of flux. For projected statistics, we start with linking the cumulants and cumulant correlators for the 
flux and the underlying neutral hydrogen density contrast $\hat \delta$. Then $\hat \delta$ is taken to be
a tracer of the underlying density contrast $\delta$, modelled statistically using the 
hierarchical ansatz. Next it is shown that, under certain simplifying assumptions, the PDF of $\delta$ 
and PDF of a suitably defined reduced flux decrement are linked through a very simple relation.  
Tests against numerical simulations shows good agreement. Results were obtained for the
entire bias functions that act as a generating function for the cumulant correlators. The scaling function $h(x)$
and the bias function $b(x)$ are also known to describe the number density and bias of over-dense objects respectively. 

The formalism developed here can also be used to probe the higher order cross-correlation statistics
involving Lyman-$\alpha$ flux decrement and weak lensing convergence of CMB maps or those extracted 
from convergence maps constructed using the weak gravitational lensing of optical galaxies.   
\section{Acknowledgements}
\label{acknow}
DM and PC acknowledge support from STFC standard grant ST/G002231/1 at
School of Physics and Astronomy at Cardiff University where this work
was completed.  MV acknowledges support from ASI/AAE, INFN PD-51, PRIN
INAF, PRIN MIUR grants and the ERC-FP7 Starting Grant ``cosmoIGM''.
We would like to thank Alan Heavens, Patrick Valageas, Ludo van
Waerbeke and Martin White for many useful discussions.  We thank
Tirthankar Roy Choudhury for his inputs.  DM would also like to thank
Francis Bernardeau for making a copy of his code available to us which
we have modified to compute the PDF and bias of the Lyman$-\alpha$
flux for the perturbative model. It is a pleasure for DM to thank
Patrick Valageas for sharing codes that were used for this study.
\bibliography{paper.bbl} \appendix

\begin{thebibliography}{}
\bibitem[\protect\citeauthoryear{Balian \& Schaeffer}{1989}]{BS89}
Balian, R., Schaeffer, 1989, A\&A, 220, 1
\bibitem[\protect\citeauthoryear{Bernardeau et al}{2002}]{Bernardreview02}
Bernardeau F., Colombi S., Gaztanaga E., Scoccimarro R., 2002, Phys.Rept.,367, 1
\bibitem[\protect\citeauthoryear{Bernardeau \& Schaeffer}{1992}]{BeS92}
Bernardeau F., Schaeffer R., 1992, A\& A, 255, 1
\bibitem[\protect\citeauthoryear{Bernardeau \& Schaeffer}{1999}]{BS99}
Bernardeau F., Schaeffer R., 1999, A\&A, 349 697
\bibitem[\protect\citeauthoryear{Bernardeau \& Kofman}{1995}]{BK95}
Bernardeau, F., Kofman, L. 1995, ApJ, 443, 479
\bibitem[\protect\citeauthoryear{Bernardeau}{1992}]{B92}
Bernardeau, F. 1992, ApJ, 392, 1
\bibitem[\protect\citeauthoryear{Bernardeau}{1994}]{B94}
Bernardeau, F. 1994, A\&A, 291, 697
\bibitem[\protect\citeauthoryear{Bernardeau \& Valageas}{2000}]{BV00} 
Bernardeau F., Valageas P., 2000, A\&A, 364, 1 
\bibitem[\protect\citeauthoryear{Bi}{1993}]{Bi93}
Bi H., 1993, ApJ, 405, 479
\bibitem[\protect\citeauthoryear{Bi \& Davidson}{1997}]{BD97}
Bi H., Davidson A.F. 1997, ApJ. 479, 523
\bibitem[\protect\citeauthoryear{Bolton, Oh, Furlanetto}{2009}]{Bolt09}
Bolton, J.S..; Oh,S.P.; Furlanetto,S.R., 2009, MNRAS, 396, 2405
\bibitem[\protect\citeauthoryear{Bolton et al.}{2005}]{Bolt05}
Bolton J.S., Haehnelt M.G., Viel M., Springle V., 2005, MNRAS, 357, 1178
\bibitem[\protect\citeauthoryear{}{}]{}
Bolton J.S., Oh S.P., Furlanetto S.R., 2009b
\bibitem[\protect\citeauthoryear{Bouchet et al}{1993}]{Bouchet93}
Bouchet, F., Strauss, M. A., Davis, M., Fisher, K. B., Yahil, A., Huchra, J. P. 1993, ApJ, 417, 36
\bibitem[\protect\citeauthoryear{Cen et al.}{1994}]{Cen94}
Cen R., Miralda-Escude' J., Ostriker J.P., Rauch M.,  1994, ApJ, 437, L83
\bibitem[\protect\citeauthoryear{Coles \& Jones}{1991}]{CJ91}
Coles P. \& Jones B. 1991, MNRAS, 248,1
\bibitem[\protect\citeauthoryear{Colombi}{1994}]{Col94}
Colombi S., 1994, ApJ, 435, L536
\bibitem[\protect\citeauthoryear{Coppolani et al.}{2006}]{Copp06}
Coppolani F., Petitjean P., Stoehr F., et al. 2006, MNRAS, 370, 1804
\bibitem[\protect\citeauthoryear{Croft et al.}{1998}]{Cr98}
Croft R.A.C., Weinberg D.H.,  katz N., Hernquist L., 1998, ApJ, 495, 44
\bibitem[\protect\citeauthoryear{Croft et al.}{1999}]{Croft99}
Croft R.A.C., Weinberg D.H., Pettini M., Hernquist L., Katz N., 1999, ApJ, 520, 1
\bibitem[\protect\citeauthoryear{Croft et al.}{2002}]{Croft02}
Croft R.A.C., Weinberg D.H., Bolte M., Burles S., Hernquist L., 
katz N., Kirkman D., Tytler D., 2002, ApJ, 581, 20
\bibitem[\protect\citeauthoryear{Croft et al.}{1999}]{Cr99}
Croft R.A.C., Weinberg D.H., Pettini M., Hernquist L., Katz N., 1999, ApJ, 520, 1
\bibitem[\protect\citeauthoryear{Davis \& Peebles}{1977}]{DP77}
Davis M., Peebles P.J.E., 1977, ApJS, 34, 425 
\bibitem[\protect\citeauthoryear{Doroshkevich \& Shandarin}{1977}]{DS97}
Doroshkevich A.G., Shandarin S.F., 1977, MNRAS, 179, 95
\bibitem[\protect\citeauthoryear{D'Odorico et al.}{2002}]{Dod02}
D'Odorico V., Petitjean P., Cristiani S., 2002, A\&A, 390, 13
\bibitem[\protect\citeauthoryear{Eisenstein \& Hu}{1998}]{EJ98}
Eisenstein D.J., Hu W., (1998), ApJ, 496, 605 
\bibitem[\protect\citeauthoryear{Fry}{1984}]{Fry84}
Fry J.N., 1984, ApJ, 279, 499
\bibitem[\protect\citeauthoryear{Gnedin \& Hui}{1996}]{GH96}
Gnedin N.Y., Hui L., 1996, ApJ, 472, L73
\bibitem[\protect\citeauthoryear{Gnedin \& Hui}{1998}]{GH98}
Gnedin N.Y., Hui L., 1998, MNRAS, 296, 44 
\bibitem[\protect\citeauthoryear{Groth \& Peebles}{1977}]{GP77}
Groth E., Peebles P.J.E., 1977, ApJ, 217,385
\bibitem[\protect\citeauthoryear{Guimaraes et al.}{2007}]{Gui07}
Guimaraes R., Petitjean P., Rollinde E. et al. 2007, MNRAS, 377, 657 
\bibitem[\protect\citeauthoryear{Gunn \& Peterson}{1965}]{GP65}
Gunn J.E. \& Peterson  B.A., 1965, ApJ,  142, 1633
\bibitem[\protect\citeauthoryear{Haardt \& Madau}{1996}]{HM96}
Haardt, F., Madau, P., 1996, ApJ, 461, 20
\bibitem[\protect\citeauthoryear{Hamilton}{1985}]{Ham85} 
Hamilton, A. J. S. 1985, ApJ, 292, L35 
\bibitem[\protect\citeauthoryear{Hui}{1999}]{Hui99}
Hui L., 1999, ApJ, 519, L9
\bibitem[\protect\citeauthoryear{Hui \& Haiman}{2003}]{HH03}
Hui L., Haiman Z., 2003, ApJ, 596, 9 
\bibitem[\protect\citeauthoryear{Hui, Gnedin \& Zhang}{1997}]{HGZ97}
Hui L.,  Gnedin Y.G., \& Zhang Y., 1997, MNRAS, 292, 27
\bibitem[\protect\citeauthoryear{Hui \& Gnedin}{1997}]{HG97}
Hui L., Gnedin N.Y., 1997, MNRAS, 292,27 
\bibitem[\protect\citeauthoryear{Kaiser}{1992}]{Kaiser92}
Kaiser N. 1992. ApJ, 388, 272
\bibitem[\protect\citeauthoryear{Kayo, Taruya, Suto}{2001}]{KTS01}
Kayo I., Taruya A., Suto Y. 2001, ApJ, 561, 22
\bibitem[\protect\citeauthoryear{Kim et al.}{2007}]{Kim07}
Kim T.-S., Bolton J.S., Viel M., Haehnelt M.G., Carswell R.F., 2007, MNRAS,
382, 1657
\bibitem[\protect\citeauthoryear{Kofman et al.}{1994}]{Kf94}
Kofman, L., Bertschinger, E., Gelb, J. M., Nusser, A.,  Dekel, A. 1994, ApJ, 420, 44 
\bibitem[\protect\citeauthoryear{Limber}{1954}]{Limb54}
Limber D.N., 1954, ApJ, 119, 665
\bibitem[\protect\citeauthoryear{Lidz et al.}{2006}]{L06}
Lidz A., Heitmann K., Hui L.,Habib S., Rauch M., Sargent W.~L.~W., 2006, ApJ, 638, 27L
\bibitem[\protect\citeauthoryear{Matarrese \& Mohayee}{2002}]{MM02}
Matarrese S.\& Mohayee R., 2002, MNRAS, 329, 37 
\bibitem[\protect\citeauthoryear{McDonald \& Miralda-Escude}{1999}]{MM99}
McDonald P., Miralda-Escude J., 1999, ApJ, 518, 24
\bibitem[\protect\citeauthoryear{McDonald et al.}{2001}]{MMR01}
McDonald P., Miralda-Escude J., Rauch M., et al. 2001, ApJ, 562, 52
\bibitem[\protect\citeauthoryear{McDonald et al.}{2005}]{Mc05}
McDonald P., Seljak U., Cen R., Bode P.,  Ostriker J.P., 2005,MNRAS, 360, 1471
\bibitem[\protect\citeauthoryear{McDonald, Seljak \& Burles.}{2006}]{MSB06}
McDonald P., Seljak U., Burles E A, 2006, ApJS, 163, 80.
\bibitem[\protect\citeauthoryear{Mcquinn et al.}{2009}]{MLZ09}
Mcquinn M., Lidz A., Zaldarriaga M., Hernquist L., Hopkins  P.F.,
Dutta S., Faucher-Giguere C.-A., 2009, ApJ, 694, 842
\bibitem[\protect\citeauthoryear{McDonald}{2003}]{Mc03}
McDonald P., 2003, ApJ, 585, 34 
\bibitem[\protect\citeauthoryear{McGill}{1990}]{McGill90}
McGill C., 1990, MNRAS, 242, 544
\bibitem[\protect\citeauthoryear{McQuinn et al.}{2009}]{Mc09}
McQuinn M., Lidz A., Zaldariagga M., Hernquist L., Hopkins P.F., Dutta S.,  
Faucher-Giguere C. -A., 2009, ApJ, 694, 842
\bibitem[\protect\citeauthoryear{McDonald \& Eisenstein}{2007}]{ME07}
McDonald P., Eisenstein D., 2007, PRD, 76, 063009
\bibitem[\protect\citeauthoryear{Meiksin \& White}{2001}]{MW01}
Meiksin A., White M., 2001, MNRAS, 324, 141
\bibitem[\protect\citeauthoryear{Munshi, Coles, Melott}{1999a}]{MCM1}
Munshi D., Coles P. Melott A., 1999, MNRAS, 310, 892
\bibitem[\protect\citeauthoryear{Munshi, Coles, Melott}{1999b}]{MCM2}
Munshi D., Coles P. Melott A., 1999, MNRAS,307, 387
\bibitem[\protect\citeauthoryear{Munshi, Melott, Coles}{1999c}]{MCM3}
Munshi D., Melott A., Coles P., 2000, MNRAS,311,149
\bibitem[\protect\citeauthoryear{Munshi \& Jain}{2000}]{MuJa00} 
Munshi D., Jain B., 2000, MNRAS, 318, 109 
\bibitem[\protect\citeauthoryear{Munshi \& Jain}{2001}]{MuJa01} 
Munshi D., Jain B., 2001, MNRAS, 322, 107 
\bibitem[\protect\citeauthoryear{Munshi \& Heavens}{2009}]{MuHe09}
Munshi D., Heavens A., MNRAS, 2010, 401, 2406 
\bibitem[\protect\citeauthoryear{Munshi et al.}{2011}]{MuShSmCo11}
Munshi D., Joudaki S., Smidt J., Coles P., 2011, MNRAS, submitted, arXiv:1106.0766 
\bibitem[\protect\citeauthoryear{Rauch et al.}{1997}]{RMS97}
Rauch M., Miralda-Escude J., Sargent W.L.W., et al., 1997, ApJ, 489, 7
\bibitem[\protect\citeauthoryear{Rauch et al.}{2005}]{Rauch05}
Rauch, M., Becker G. D., Viel M., Sargent W. L. W., Smette, A., Simcoe R. A. 
Barlow T. A., Haehnelt, M. G., 2005, ApJ, 632, 58
\bibitem[\protect\citeauthoryear{Ricotti, Gnedin \& Shull}{2000}]{RGS00}
Ricotti M., Gnedin N.Y., Shull J.M., 2000, ApJ, 534, 41
\bibitem[\protect\citeauthoryear{Rolinde et al.}{2003}]{Rol03}
Rollinde E., Pettijean P., Pichon  C. et al., 2003, MNRAS, 341, 1279
\bibitem[\protect\citeauthoryear{Roy Choudhury, Padamanabhan \& Srianand}{2001}]{RPT01}
Roy Choudhury T., Padmanabhan T., Srianand R., MNRAS, 2001, 322, 561
\bibitem[\protect\citeauthoryear{Saitta et al.}{2008}]{Saitta08}
Saitta F., D'Odorico V., Bruscoli M., Cristiani S., Monaco P., Viel M., 2008, MNRAS, 385, 519
\bibitem[\protect\citeauthoryear{Schaye et al.}{1999}]{STLE99}
Schaye J., Theuns T., Leonard A., Efstathiou G., 1999, MNRAS, 310, 57 
\bibitem[\protect\citeauthoryear{Seljak, Slosar, McDonald}{2006}]{SSM06}
Seljak U., Slosar A., McDonald P., 2006, JCAP, 10,14
\bibitem[\protect\citeauthoryear{Slosar et al.}{2011}]{Sls11}
Slosar et al., 2011, JCAP, 09, 001
\bibitem[\protect\citeauthoryear{Springel et al.}{2005}]{Sp05}
Springel V., 2005, MNRAS, 364, 1105
\bibitem[\protect\citeauthoryear{Szapudi \& Szalay}{1993}]{szsz93} 
Szapudi I., Szalay A.S., 1993, ApJ, 408, 43
\bibitem[\protect\citeauthoryear{Szapudi \& Szalay}{1997}]{szsz97} 
Szapudi I., Szalay A.S., 1997, ApJ, 481, L1
\bibitem[\protect\citeauthoryear{Szapudi \& Szalay}{1997}]{SS97}
Szapudi I., Szalay A.S., 1999, ApJ, 515, L43 
\bibitem[\protect\citeauthoryear{Taruya et al.}{2002}]{Taruya02}
Taruya A., Takada M., Hamana T., Kayo I., Futamase T., 2002, ApJ, 571, 638 
\bibitem[\protect\citeauthoryear{Theuns et al.}{1998}]{TLEPT98}
Theuns T., Leonard A., Efstathiou G., Pearce F.R., Thomas P.A., 1998, MNRAS, 301, 478
\bibitem[\protect\citeauthoryear{Theuns et al.}{2002}]{TSS02}
Theuns T., Schaye J., Zaroubi S.. et al. 2002, ApJ, 567, L103
\bibitem[\protect\citeauthoryear{Theuns et al.}{2002}]{T02}
Theuns T., Viel M., Kay S.. et al. 2002, ApJ, 578, L5
\bibitem[\protect\citeauthoryear{Tytler et al}{2004}]{T04}
Tytler et al. 2004, ApJ,617,1
\bibitem[\protect\citeauthoryear{Valageas \& Munshi}{2004}]{VaMu04}
Valageas P., Munshi D., 2004, MNRAS, 354, 1146 
\bibitem[\protect\citeauthoryear{Valageas}{2000}]{V00}
Valageas P., 2000, A\&A, 354,767
\bibitem[\protect\citeauthoryear{Valageas, Schaeffer \& Silk}{1999}]{VSS99}
Valageas P., Schaeffer R., Silk J., 1999, A\&A, 345, 691 
\bibitem[\protect\citeauthoryear{Valageas \& Schaeffer}{2000}]{VS00}
Valageas P., Schaeffer R., 2000, A\&A, 356, 771 
%
\bibitem[\protect\citeauthoryear{Valageas, Silk \& Schaeffer}{2001}]{VSS01}
Valageas P., Silk J.,Schaeffer R., A\&S. 2001, 366, 363
\bibitem[\protect\citeauthoryear{Valageas, Schaeffer \& Silk}{2002}]{VSS02}
Valageas P., Schaeffer R., Silk J., 2002, A\&A, 388, 741
\bibitem[\protect\citeauthoryear{Valageas \& Silk}{1999}]{VS99a}
Valageas P., Silk J., 1999, A\&A, 347, 1 
\bibitem[\protect\citeauthoryear{Valageas \& Schaeffer}{1999}]{VS99}
Valageas P., Schaeffer R., 1999, A\&S, 345, 329
\bibitem[\protect\citeauthoryear{Valageas, Balbi \& Silk}{2001}]{VS01}
Valageas P., Balbi A., Silk J., A\&A, 2001, 367, 1 
\bibitem[\protect\citeauthoryear{Valageas \& Schaeffer}{2000}]{VS00b}
Valageas P., Schaeffer R., 2000, A\&A, 359, 821 
\bibitem[\protect\citeauthoryear{Valageas}{2001}]{Va00}
Valageas P., 2000, A\&A, 354, 767
\bibitem[\protect\citeauthoryear{Valentino et al.}{2011}]{VVDS11}
Vallinotto A., Viel M., Das S., Spergel D.N., 2011, ApJ, 735, 38 
\bibitem[\protect\citeauthoryear{Vallinotto et al.}{2009}]{VDSS09}
Vallinotto A., Das S., Spergel D.N., Viel M., 2009, PRL, 103, 091304 
\bibitem[\protect\citeauthoryear{Viel et al.}{2002}]{Viel02}
Viel M., Matarrese S., Mo H.J., Haehnelt M.G., Thuns T., 2002, MNRAS, 329, 848
\bibitem[\protect\citeauthoryear{Viel \& Haehnelt}{2006}]{VH06}
Viel M., Haenelt M.G., 2006, MNRAS, 365, 231
\bibitem[\protect\citeauthoryear{Viel, Haenelt \& Springel}{2004}]{VHS04}
Viel M., Haenelt M.G., Springel V., 2004, MNRAS, 354, 684
\bibitem[\protect\citeauthoryear{Viel et al}{2008}]{Viel08}
Viel M., Becker G.D., Bolton J.S., Haehnelt M.G., Rauch M., Sargent W.L.W., 2008, PRL, 100, 041304
\bibitem[\protect\citeauthoryear{Viel et al.}{2009}]{Viel09}
Viel M., Branchini E., Dolag K., Grossi M., Matarrese S., Moscardini L., 2009, MNRAS, 393, 774
\bibitem[\protect\citeauthoryear{Zel'dovich}{1970}]{Ze70}
Zel'dovich ya. B., 1970, A\&A, 5, 84
\end{thebibliography}
\section{Hierarchical Ansatz: A Very Brief Review}
\label{hier}
There is no complete analytical model for the evolution of gravitational clustering in the nonlinear regime and the
Eulerian and Lagrangian perturbation theories \citep{Bernardreview02} are often used to describe clustering in the quasi-linear regime. 
The halo model is extensively used and is very successful in modeling fully evolved structure formation. A parallel 
approach, depends on the hierarchical nature of the correlation functions 
that develop during the nonlinear regime.It has a long history in describing clustering dark matter distribution
as well as the collapsed objects. Additional assumptions regarding virialization and hydrodynamical equilibrium
can be used in association to make specific predictions regarding diverse cosmological phenomenon from Luminosity 
of X-ray clusters \citep{VS00b}, column density distribution of neutral hydrogen \citep{VSS02}, 
redshift evolution bias \citep{VSS01}, mass and luminosity distribution of galaxies and clusters \citep{VS99},
cosmic microwave background (CMB) secondaries such as the thermal and kinetic Sunyaev and Zel'dovich effect \citep{MuShSmCo11,VS01},
reheating and reionization of the Universe \citep{VS99,VSS02}. Weak lensing observables have already
been studied in this framework \citep{Va00,BV00,MuJa00,MuJa01,VS99}.

The hierarchical ansatz \citep{BS89} is remarkably successful at making approximate calculations of the entire PDF and bias of the
density field, improving significantly upon the order-by-order analysis of other approaches (\cite{Bernardreview02} for a detailed
review). The hierarchical ansatz
in the highly non-linear regime depends on assuming a specific correlation hierarchy, on the other hand
in the quasi-linear regime it can be linked to the gravitational dynamics using perturbative analysis. 
\subsection{Highly Non-linear Regime}
The PDF $p(\delta)$ and the bias $b(\delta)$ can both be constructed from the knowledge of the VPF $\phi(y)=\sum_{p=1} S_p {y^p/p!}$
and its two-point analog $\tau(y) =\sum_{p}C_{p1} {y^p/p!}$. Where the parameters $S_p$ and $C_{p1}$ are normalized 
cumulants and cumulant correlators for the density field. The PDF is related to the scaling function $h(x)$ in the text of the
paper $p(\delta)= h(x)/\bar \xi_2^2$ with $x={(1+\delta)/\bar \xi_2}$.
\be
p(\delta) = \int_{-i\infty}^{i\infty} { d{y} \over 2 \pi i} \exp \Big [ {(
1 + \delta )y - \phi({y})  \over \bar \xi_2} \Big ]; \quad
b(\delta) p(\delta) = \int_{-i\infty}^{i\infty} { dy \over 2 \pi i} \tau(y) \exp \Big [ {(
1 + \delta )y - \phi(y)  \over \bar \xi_2} \Big ]
\label{eq:ber}.
\ee
The modelling of $\phi(y)$ and $\tau(y)$ needs a detailed knowledge of 
the entire correlation hierarchy. The detailed knowledge of the entire correlation hierarchy is
encoded in the vertex generating function ${\cal G}(\tau))$. Typically for large values of
$y$ the VPF exhibits a power law $\phi(y)=ay^{1-\omega}$. There are no theoretical estimates of $\omega$ and it is
generally estimated from numerical simulations. The parameter typically takes a value
$\omega=0.3$ for CDM like spectra. For small but negative values of $y$ 
the functions $\phi(y)$ and $\tau(y)$ develops a singularity in the complex plane which
is described by the following parametrization.
\be
\phi(y) = \phi_s - a_s \Gamma(\omega_s) ( y - y_s)^{-\omega_s}; \quad\quad
\tau(y) = \tau_s - b_s ( y - y_s )^{-\omega_s - 1}.
\ee
The singularity structure of $\phi(y)$ and $\tau(y)$ depends on the nature of the vertex generating
function $G(\tau)$ and its behaviour near the singularity $\tau_s$:
\be
a_s = {1 \over \Gamma(-1/2)}{\cal G}'(\tau_s) {\cal G}''(\tau_s) \left [
{ 2 {\cal G}'(\tau_s) {\cal G}''(\tau_s) \over {\cal G}'''(\tau_s)}
\right ]^{3/2}; \quad\quad
b_s = \left [
{ 2 {\cal G}'(\tau_s) {\cal G}''(\tau_s) \over {\cal G}'''(\tau_s)}
\right ]^{1/2}.
\ee
On the other hand the parameters $\omega$ and $a$ can be described in terms of a parameter $y_s$
which in turn describes the exponential decay of the PDF for large density contrast $\delta$:
\be
\omega = k_a / ( k_a + 2),\label{ka}; \quad\quad
a = {k_a + 2 \over 2} k_a^{ k_a /  k_a + 2}; \quad\quad
-{ 1 \over y_s} = x_{\star} = {1 \over k_a } { (k_a + 2)^{k_a + 2} \over (k_a + 1)^{k_a+1}}.
\ee
The PDF and the bias thus has two distinct regimes that are dictated by the two asymptotes.
For intermediate values of $\delta$ the PDF shows a power law behaviour. The PDF
and the bias are given by the following expression:
\begin{equation}
{\bar \xi }^{ - \omega \over ( 1 - \omega)} << 1 + \delta << \bar \xi;
~~~~~~
p(\delta) = { a \over \bar \xi_2^2} { 1- \omega \over \Gamma(\omega)}
\Big ( { 1 + \delta \over \xi_2 } \Big )^{\omega - 2}; ~~~~~
 b(\delta) = \left ( {\omega \over 2a } \right )^{1/2} { \Gamma
(\omega) \over \Gamma [ { 1\over 2} ( 1 + \omega ) ] } \left( { 1 +
\delta \over \bar \xi_2} \right)^{(1 - \omega)/2}.
\end{equation}
For large values of $\delta$ the PDF on the other hand shows an exponential
behaviour:
\begin{equation}
1+ \delta >> {\bar \xi}_2; ~~~~
p(\delta) = { a_s \over \bar \xi_2^2 } \Big ( { 1 + \delta \over \bar
\xi_2}  \Big ) \exp \Big ( - { 1 + \delta \over x_{\star} \bar \xi_2}
\Big );  ~~~~~ b(\delta) = -{ 1 \over {\cal G}'(\tau_s)} {(1 + \delta)
\over { {\bar \xi}_2}}.
\end{equation}
At very small values of $\delta$ the PDF shows an exponential decay which
is described only by the parameter $\omega$:
\begin{equation}
1 + \delta << \bar \xi_2;~~~~
p(\delta) = a^{ -1/(1 - \omega)} {\bar \xi}_2^{ \omega/( 1 -
\omega )} \sqrt { ( 1 - \omega )^{ 1/\omega } \over 2 \pi \omega z^{(1
+ \omega)/ \omega } } \exp \Big [ - \omega \Big ( {z \over 1 - \omega}
\Big )^{- {{(1 - \omega)}/\omega}} \Big ]; ~~~~~~~b(\delta) = -
\left ( {2 \omega \over \bar{ \xi}_2} \right )^{1/2} \left ({ 1 -
\omega \over z}  \right )^{(1 - \omega)/2 \omega}.
\end{equation}
The range of $\delta$ for which the power law regime is valid depends on 
the value of $\bar\xi_2$. For smaller values of $\bar\xi_2$ the power law 
regime is less pronounced.
\subsection{Quasi-linear Regime}
In the quasi-linear regime the parameters that describe the generating functions can be
directly lniked with gravitational dynamics \citep{B92,B94}. The PDF and bias in the intermediate power-law regime
in the quasi-linear regime is given by the following result:
\begin{eqnarray}
&&p(\delta)d \delta = { 1 \over -{\cal G}_{\delta}'(\tau) } \Big [ { 1 - \tau {\cal G}_{\delta}''(\tau)
/{\cal G}_{\delta}'(\tau) \over 2 \pi {\bar \xi}_{2} }  \Big ]^{1/2} \exp \Big ( -{ \tau^2
\over 2 {\bar \xi}_{2}} \Big ) d \tau; ~~~~~ b(\delta) = - \left (
{k_a \over \bar \xi_2} \right ) \left [ ( 1 + {\cal G}_{\delta}(\tau)
)^{1/k_a} - 1 \right ]; \\
&&{\cal G}_{\delta}(\tau) = {\cal G}(\tau) - 1 =  \delta.
\end{eqnarray}
For the large density contrast the exponential decay take the following form:
\begin{equation}
p(\delta) d \delta = { 3 a_s \sqrt {{\bar \xi}_2} \over 4  {\sqrt
\pi} } \delta^{-5/2} \exp \Big [ -|y_s|{ \delta \over {\bar
\xi}_{2}} + {|\phi_s| \over {\bar \xi}_{2}} \Big ] d \delta;
~~~~~b(\delta) = -{ 1 \over {\cal G}'(\tau_s)} {(1 + \delta) \over {
{\bar \xi}_2}}.
\end{equation}
These results ignore the loop corrections to tree-level perturbation theory but 
take into account correlation hierarchy to an arbitrary order. 
\section{Lognormal Distribution}
\label{logn}
The evolution of the PDF $p(\delta)$ of density field $\delta$ due to gravitational clustering has been studied
extensively in many cosmological context. The lognormal distribution \citep{Ham85,CJ91,Bouchet93,Kf94} is known to accurately reproduce the results
from numerical simulation in the quasi-linear regime and has been used to model both one- and two-point PDF. In general 
the lognormal distribution can provide a good statistical description of a random variable if it can be modeled as
a product of many other random variables \citep{KTS01,Taruya02}.
\begin{eqnarray}
&& p(\delta)d\delta = {1 \over \sqrt {2\pi \Sigma}} \exp \left [ -{\Lambda^2 \over 2\Sigma}\right ]{d\delta \over 1+\delta};
\label{eq:logn1}\\
&& \Sigma_{}=\ln(1+\sigma^2); \quad \Lambda = \ln[(1+\delta)\sqrt{(1+\sigma^2)};\label{eq:logn1a} \\
&& p(\delta_1,\delta_2)d\delta_1 d\delta_2 =  {1 \over 2\pi \sqrt {\Sigma^2 - X_{12}^2}}\exp \left [ -{\Sigma(\Lambda_1^2 + \Lambda_2^2) -2X_{12}\Lambda_1\Lambda_2 \over 2(\Sigma^2 - X_{12}^2)}\right ] {d\delta_1\over 1+\delta_1} {d\delta_2\over 1+\delta_2}; \label{eq:logn2}\\
&& \Lambda_i = \ln[(1+\delta_i)\sqrt{(1+\sigma^2)};  \quad X_{12}=\ln(1+\xi_{12})
\label{eq:logn2a}
\end{eqnarray}
In the lowest order in $\xi_{12}$ the lognormal model takes a factorized form. In this limit a 
bias function $b(\delta)$ can be defined.
\be
p(\delta_1,\delta_2)= p(\delta_1)p(\delta_2)[1+ b(\delta_1)\xi_{12}b(\delta_2)]; \quad\quad  b(\delta_i)= \Lambda_i/\Sigma_{}.
\label{eq:log_bias}
\ee
The lognormal distribution has also been used to model many cosmological observation from galaxy surveys to weak lensing
data.
\end{document}